\documentclass[english,12pt,a4paper,english,pre,onecolumn,superscriptaddress,floatfix]{revtex4-1}
\usepackage[T1]{fontenc}
\usepackage[latin9]{inputenc}
\usepackage{color}
\usepackage{babel}
\usepackage{textcomp}
\usepackage{mathrsfs}
\usepackage{algorithm2e}
\usepackage{amsmath}
\usepackage{amssymb}
\usepackage{graphicx}
\usepackage{esint}
\usepackage[unicode=true,pdfusetitle,
 bookmarks=true,bookmarksnumbered=false,bookmarksopen=false,
 breaklinks=false,pdfborder={0 0 0},backref=false,colorlinks=true]
 {hyperref}
\hypersetup{
 linkcolor=coolblack,citecolor=burntorange}

\makeatletter

\newcommand{\lyxdot}{.}

\usepackage{mathrsfs}
\DeclareMathOperator*{\sign}{sign}
\DeclareMathOperator*{\argmax}{argmax}

\LinesNumbered
\RestyleAlgo{algoruled}
\DontPrintSemicolon

\definecolor{burntorange}{rgb}{0.8, 0.33, 0.0}
\definecolor{charcoal}{rgb}{0.21, 0.27, 0.31}
\definecolor{coolblack}{rgb}{0.0, 0.18, 0.39}

\makeatother

\begin{document}

\title{Local entropy as a measure for sampling solutions in Constraint Satisfaction
Problems}

\author{Carlo Baldassi}

\author{Alessandro Ingrosso}

\author{Carlo Lucibello}

\author{Luca Saglietti}

\affiliation{Dept. Applied Science and Technology, Politecnico di Torino, Corso
Duca degli Abruzzi 24, I-10129 Torino, Italy}

\affiliation{Human Genetics Foundation-Torino,  Via Nizza 52, I-10126 Torino,
Italy}

\author{Riccardo Zecchina}

\affiliation{Dept. Applied Science and Technology, Politecnico di Torino, Corso
Duca degli Abruzzi 24, I-10129 Torino, Italy}

\affiliation{Human Genetics Foundation-Torino, Via Nizza 52, I-10126 Torino, Italy}

\affiliation{Collegio Carlo Alberto, Via Real Collegio 30, I-10024 Moncalieri,
Italy}
\begin{abstract}
We introduce a novel Entropy-driven Monte Carlo (EdMC) strategy to
efficiently sample solutions of random Constraint Satisfaction Problems
(CSPs). First, we extend a recent result that, using a large-deviation
analysis, shows that the geometry of the space of solutions of the
Binary Perceptron Learning Problem (a prototypical CSP), contains
regions of very high-density of solutions. Despite being sub-dominant,
these regions can be found by optimizing a local entropy measure.
Building on these results, we construct a fast solver that relies
exclusively on a local entropy estimate, and can be applied to general
CSPs. We describe its performance not only for the Perceptron Learning
Problem but also for the random $K$-Satisfiabilty Problem (another
prototypical CSP with a radically different structure), and show numerically
that a simple zero-temperature Metropolis search in the smooth local
entropy landscape can reach sub-dominant clusters of optimal solutions
in a small number of steps, while standard Simulated Annealing either
requires extremely long cooling procedures or just fails. We also
discuss how the EdMC can heuristically be made even more efficient
for the cases we studied.

\tableofcontents{}
\end{abstract}
\maketitle

\section{Introduction\label{sec:Introduction}}

Markov Chain Monte Carlo (MCMC) algorithms for combinatorial optimization
problems are designed to converge to a stationary distribution $\pi$
that is a monotone decreasing function of the objective function one
needs to minimize. A fictitious temperature is usually introduced
to make the distribution more and more focused on the optima. Depending
on the form of the stationary distribution (i.e. on the temperature)
the sampling process can converge fast or it can get trapped in local
minima. There is typically a tradeoff between optimality of the sampled
solutions and the form of $\pi$: smooth and close to uniform distributions
are the easiest to sample but the sampled configuration are most often
far from optimal. On the contrary hard to sample distributions are
characterized by a so called glassy landscape where the number of
metastable minima that can trap the MCMC and brake ergodicity are
typically exponentially numerous \cite{krzakala2007gibbs,mezard_information_2009}. 

Random constraint satisfaction problems (CSPs) offer an ideal framework
for understanding these type of questions in that they allow to analyze
the geometry of the space solutions of hard to sample problems and
at the same time design and test novel algorithms \cite{mezard2002analytic}.
In many random CSPs, the computational hardness is associated to the
existence of optimal and metastable states that are grouped into different
clusters of nearby solutions with different sizes and properties.
Finer geometrical properties of the space of solutions have been investigated
in the literature \cite{Zhou2009,Li2009,Mezard2005}, but a general
scenario has yet to be established.

Large deviation analysis allows to describe in some detail the structures
of such clusters, ranging from the dominant clusters (those in which
one would fall by choosing uniformly at random a solution) to the
subdominant ones. Note that in general, algorithms are by no means
bounded to sample solutions uniformly at random.

Very recently it has been shown \cite{baldassi-subdominant} that
problems that were believed to be intractable due to their glassy
nature, namely the learning problems in neural networks with discrete
synaptic weights, possess in fact a richer structure of the space
of solutions than what is suggested by the standard equilibrium analysis.
As it turns out, there exist sub-dominant and extremely dense clusters
of solutions that can be analytically unveiled by defining a probabilistic
weight based on the local entropy, i.e.~on the number of solutions
within a given radius from a reference solution. Despite being sub-dominant,
these states turn out to be accessible by extremely simple heuristic
algorithms.

Here, we move forward and study the same structures without enforcing
the constraint that the reference configuration is itself a solution.
This apparent simplification actually requires higher levels of replica-symmetry
breaking in the analysis. The resulting measure defines an objective
function, namely the local entropy, that we then use to devise a novel
MCMC, which we call Entropy-driven Monte Carlo (EdMC). When applied
to the binary perceptron learning problem, EdMC yields algorithmic
results that are in very good agreement with the theoretical computations,
and by far outperform standard Simulated Annealing in a direct comparison.

We also applied this approach to the $K$-SAT problem, showing that
even when some of the computations can not be carried out exactly
the practical performance is indeed very good even in hard regions.
Further heuristic algorithmic improvements are also discussed.

Here, we focus on the representative case of zero temperature and
Hamming distance, but the technique could be easily extended to go
beyond these simple assumptions.

The rest of the paper is organized as follows: in Sec.~\ref{sec:Sub-dominant-clusters-analysis}
we present the results of the analysis of the sub-dominant clusters
of solutions and define the novel EdMC algorithm, in Sec.~\ref{sec:EdMC-results}
we report extensive numerical results, comparing EdMC to the theoretical
results and to standard Simulated Annealing, while in Sec.~\ref{sec:Discussion}
we discuss our findings. Two Appendices follow the main text: in the
first, Appendix~\ref{sec:APPENDIX-I-BP}, we provide details about
the Belief Propagation algorithm; in the second, Appendix~\ref{sec:APPENDIX-II-Details},
we report a detailed self-contained description of the analytical
computations.

\subsection{Sub-dominant clusters analysis and Entropy-driven Monte Carlo\label{sec:Sub-dominant-clusters-analysis} }

In most random combinatorial optimization problems the so called dominating
states of the equilibrium Gibbs measure at zero temperature (the ground
states) are not relevant in the analysis of practical optimization
algorithm and their dynamics. The algorithmically accessible states
are typically sub-dominant states characterized by a high internal
entropy \cite{dall2008entropy}. The structure of such sub-dominant
states can be investigated by means of the Replica Method or the Cavity
Method, at least in the average case.

The approach we will present in the following extends our recent findings
in the discrete Perceptron Learning Problem \cite{baldassi-subdominant},
where we carried out a large deviations analysis by introducing a
reweighting of the solutions by the \emph{local entropy}, i.e.~by
the number of other solutions surrounding a given one. We show that
even if we remove the explicit constraint that the reference configuration
is a solution, optimizing the local entropy has approximately the
same effect with high probability: in other words we find that if
we can estimate the local entropy for any given configuration, and
then seek the configuration for which it is maximum, in all likelihood
we will end up in a solution.

This naturally raises the possibility to translate such theoretical
predictions into a general practical solver, because even though computing
the local entropy may be more difficult than computing the energy,
the resulting landscape may be radically different. A simple and direct
way to show that this is the case is to compare two MCMC implementations,
one optimizing the energy and one optimizing the local entropy (we
call the latter Entropy-driven Monte Carlo, or EdMC). Indeed, the
local entropy landscape proves to be much smoother, allowing to reach
ground states inaccessible to the energetic MCMC.

\subsubsection{Large deviations analysis of CSPs}

A generic Constraint Satisfaction Problem (CSP) can be defined in
terms of configurations of $N$ variables $x_{i}\in X_{i}$, subject
to $M$ constraints $\psi_{\mu}:\,D_{\mu}\to\left\{ 0,1\right\} $.
Each constraint $\mu$ involves a subset $\partial\mu$ of the variables,
which we collectively represent as $x_{\partial\mu}=\left\{ x_{i}:\,i\in\partial\mu\right\} \in D_{\mu}$,
and we define $\psi_{\mu}\left(x_{\partial\mu}\right)=1$ if the constraint
is satisfied, $0$ otherwise. For concreteness, let us focus on the
case of binary spin variables $X_{i}=X=\left\{ -1,+1\right\} $, the
generalization to multi-valued variables being straightforward \cite{baldassi-subdominant-multivalued}.
We may define an energy function of the system simply as the number
of violated constraints, namely:

\begin{equation}
H\left(x\right)=\sum_{\mu}E_{\mu}\left(x_{\partial\mu}\right)=\sum_{\mu}\left(1-\psi_{\mu}\left(x_{\partial\mu}\right)\right)\label{eq:hamiltonian_generic}
\end{equation}

A solution of a CSP is then a zero-energy configuration. The standard
zero-temperature Gibbs measure for CSPs, which we will call equilibrium
measure, assumes uniform weight over the space of solutions and it
is the one associated to the partition function

\begin{equation}
\mathcal{N}=\lim_{\beta\to\infty}\sum_{x}e^{-\beta H(x)}=\sum_{x}\prod_{\mu}\psi_{\mu}\left(x_{\partial\mu}\right)\label{eq:num_sol}
\end{equation}

which just counts the solutions.

Suppose now one wants to analyze the local structure of the solution
space by counting the number of solution vectors $x$ around a given
\emph{planted} vector $\tilde{x}$. To this end, we define the \emph{local
free entropy}, as a function of the planted configuration, as:
\begin{equation}
F\left(\tilde{x},\gamma\right)=\frac{1}{N}\log\mathcal{N}\left(\tilde{x},\gamma\right)\label{eq:local_free_entropy}
\end{equation}
where
\begin{equation}
\mathcal{N}(\tilde{x},\gamma)=\sum_{x}\prod_{\mu}\psi_{\mu}\left(x_{\partial\mu}\right)e^{\gamma x\cdot\tilde{x}}\label{eq:local_num_sol_soft}
\end{equation}
counts all solutions to the CSP with a weight that depends on the
distance from $\tilde{x}$ and is modulated by the parameter $\gamma$.
Solving a CSP amounts to finding a solution vector $\tilde{x}^{\star}$
such that $H\left(\tilde{x}^{\star}\right)=0$: we shall show below
that this can be achieved by optimizing the cost function $F\left(\tilde{x},\gamma\right)$
over $\tilde{x}$, which naturally guides the system in a region with
a high density of solutions. In order to determine $F\left(\tilde{x},\gamma\right)$,
one needs to study a slightly different system then the one defined
by $H\left(x\right)$, in which the variables $x_{i}$'s are coupled
to some external fields $\gamma\tilde{x_{i}}$, with $\tilde{x}_{i}\in\left\{ -1,+1\right\} $
and $\gamma\in\mathbb{R}$ is the coupling strength. Thus the directions
of the external fields $\tilde{x}_{i}$'s are considered as external
control variables. The parameter $\gamma$ sets the magnitude of the
external fields, and in the limit of large $N$ it effectively fixes
the Hamming distance $d$ of the solutions $x$ from $\tilde{x}$.
The local free entropy $F\left(\tilde{x},\gamma\right)$ is then obtained
as the zero-temperature limit of the free energy of the system described
by $H\left(x;\tilde{x}\right)$, and can be computed by Belief Propagation
(see Sec.~\ref{sub:EdMC:-Local-entropy} and Appendix~\ref{sec:APPENDIX-I-BP}).

We then study a system defined by the following free energy: 
\begin{equation}
\mathscr{F}\left(\gamma,y\right)=-\frac{1}{Ny}\log\left(\sum_{\tilde{x}}e{}^{yNF\left(\tilde{x},\gamma\right)}\right)\label{eq:FreeEnergy}
\end{equation}
where $y$ has formally the role of an inverse temperature and $-F\left(\tilde{x},\gamma\right)$
has formally the role of an energy. Throughout the paper the term
temperature will always refer to \textbf{$y^{-1}$} except where otherwise
stated. In the limit of large $y$, this system is dominated by the
ground states $\tilde{x}^{\star}$ for which the local free entropy
is maximum; if the number of such ground states is not exponentially
large in $N$, the \emph{local entropy} can then be recovered by computing
the Legendre transform
\begin{equation}
\mathscr{S}\left(\gamma,\infty\right)=-\mathscr{F}\left(\gamma,\infty\right)-\gamma S\label{eq:local_entropy}
\end{equation}
where $S$ is the typical value of the overlap $\frac{x\cdot\tilde{x}^{\star}}{N}$.
This quantity thus allows us to compute 
\begin{equation}
e^{N\mathscr{S}\left(\gamma,\infty\right)}=\mathcal{N}\left(\tilde{x}^{\star},\gamma\right)=\sum_{x}\prod_{\mu}\psi_{\mu}\left(x_{\partial\mu}\right)\delta\left(NS-x\cdot\tilde{x}^{\star}\right)
\end{equation}
 i.e.~to count the number of solutions at normalized Hamming distance
$d=\frac{1-S}{2}$ from the ground states $\tilde{x}^{\star}$ (note
that the soft constraint of eq.~\ref{eq:local_num_sol_soft} is equivalent
to a hard constraint in the limit of large $N$, but that the opposite
is not true in general --- see Sec.~\ref{sec:APPENDIX-II-Intro}
for more details on this point).

Informally speaking, if the distance $d$ is small enough (i.e.~at
large $\gamma$), then $\tilde{x}^{\star}$ is going to be roughly
at the center of a dense cluster of solutions, if such cluster exists,
which means that it is likely going to be a solution itself (indeed,
that is surely the case when $d=0$). Furthermore, because of the
reweighting term, these dense solution regions will typically have
different statistical properties with respect to the set of thermodynamically
locally stable states. We therefore refer to these solutions as \emph{sub-dominant
clusters}, since they are not normally part of the equilibrium description.
However, we have reasons to believe that these kind of sub-dominant
clusters play a crucial role in the algorithmic properties of practical
learning problems in CSPs: the analysis of the local free entropy
$F\left(\tilde{x}^{\star},\gamma\right)$ in the case of the binary
perceptron learning problem (see below) has shown that indeed such
sub-dominant ultra-dense regions exist at least up to a critical value
of the parameter $\alpha=\frac{M}{N}$, that these kind of solutions
exhibit different properties with respect to typical equilibrium solutions,
and that heuristic solvers typically find a solution in such regions
\cite{baldassi-subdominant}.

\subsubsection{Two prototypical CSPs\label{sub:Two-prototypical-CSPs}}

In this paper we consider two prototypical CSPs, which are both computationally
hard but have very different characteristics and structure. The first
one --- the main focus of this paper --- is the binary Perceptron
Learning Problem, a fully connected problem that originally motivated
our investigation. The second is an example of a general diluted problem,
the Random $K$-SAT, which has a long tradition in the Statistical
Mechanics of Optimization Problems.

\paragraph{Binary perceptron\label{par:Binary-perceptron}}

Let us consider the problem of classifying $M=\alpha N$ input patterns
$\xi^{\mu}\in\left\{ -1,+1\right\} ^{N}$, that is associating to
each of them a prescribed output $\sigma^{\mu}\in\left\{ -1,+1\right\} $.
The perceptron is defined as a simple linear threshold unit that implements
the mapping $\tau\left(x,\xi\right)=\mathrm{sign}\left(x\cdot\xi\right)$,
with $x$ representing the vector of \emph{synaptic weights}. In what
follows we will consider the classification problem, which consists
in finding a vector $x^{\star}$ that correctly classifies all inputs,
i.e. $\tau(x^{\star},\xi^{\mu})=\sigma^{\mu},\;\mu\in\left\{ 1,...,M\right\} $,
given a set of random i.i.d. unbiased $\left\{ \xi_{i}^{\mu},\sigma^{\mu}\right\} $.
We will focus on the case of spin-like weights, i.e. $x\in\left\{ -1,+1\right\} $
(the generalization to more states does not pose significant additional
difficulties). The corresponding energy function is the sum of wrongly
classified patterns, namely:
\begin{equation}
H_{\mathrm{perc}}\left(x\right)=\sum_{\mu}\Theta\left(-\sigma^{\mu}\tau\left(x,\xi^{\mu}\right)\right)\label{eq:hamiltonian_perc}
\end{equation}
where $\Theta\left(\cdot\right)$ is the Heaviside step function.
This problem has been extensively analyzed in the limit of large $N$
by means of Replica \cite{gardner-derrida,krauth-mezard} and Cavity
\cite{mezard-gardner-cavity} methods, finding that in the typical
case there is an exponential (in $N$) number of solutions up to a
critical capacity $\alpha_{c}=0.833$, above which no solution typically
exists. Despite the exponential number of solutions, the energetic
landscape is riddled with local minima, and energetic local search
algorithms are typically ineffective at finding them for any $\alpha$
\cite{obuchi2009weight,huang2013entropy,huang2014origin}. Moreover
typical solutions are known to be isolated \cite{huang2014origin}.

\paragraph{$K$-SAT}

The satisfiability problem, in its `random $K$-SAT' instantiation,
consists in finding an assignment for $N$ truth values that satisfies
$M=\alpha N$ random logical clauses, each one involving exactly $K$
different variables. Let us then consider $N$ Boolean variables $\left\{ t_{i}\in\left\{ 0,1\right\} \right\} $,
with the common identification $\left\{ 0\to\mathrm{FALSE},\;1\to\mathrm{TRUE}\right\} $.
A given clause $\mu$ is the logical OR of its variables, whose indices
are $i_{1}^{\mu},...,i_{K}^{\mu}$, and which can appear negated.
Let us work in a spin representation $x_{i}=2t_{i}-1$ and introduce
the couplings $J_{i_{r}}^{\mu}\in\left\{ -1,+1\right\} $, where $J_{i_{r}}^{\mu}=1$
if the variable $x_{i_{r}^{\mu}}$ appears negated in clause $\mu$,
and $J_{i_{r}}^{\mu}=-1$ otherwise. The graph structure is random,
in that each clause involves $K$ variables extracted uniformly at
random, and the couplings are also unbiased i.i.d.~random binary
variables. With these definitions, the solutions to a $K$-SAT problem
are the zero energy configurations of the following Hamiltonian:
\begin{equation}
H_{\mathrm{SAT}}=2\sum_{\mu=1}^{M}\prod_{k=1}^{K}\left(\frac{1+J_{i_{r}}^{\mu}x_{i_{r}^{\mu}}}{2}\right)
\end{equation}
which counts the number of violated clauses.

Random $K$-SAT has been central in the development of the statistical
mechanical approach to CSPs. For more details, we refer the reader
to comprehensive reviews \cite{martin2001statistical,mezard_information_2009}.
Here we just point out that, when varying the number of constraints
per variable $\alpha$, the problem undergoes a sequence of phase
transitions, related to the fragmentation of the phase space in a
huge number of disconnected clusters of solutions. This rich phenomenology,
observed well below the UNSAT threshold (above which no solution typically
exists at large $N$), can be analyzed by the cavity method in the
framework of $1$-step Replica Symmetry Breaking ($1$-RSB), and is
reflected in the exponential slowing down of greedy algorithms as
well as sampling strategies.

\subsubsection{1-Step Replica-Symmetry-Broken solution in the binary perceptron\label{sub:1RSB}}

The existence of dense sub-dominant regions of solutions with radically
different properties than those described by the usual equilibrium
analysis was first discovered for the perceptron learning problem
in \cite{baldassi-subdominant}, as a result of extensive numerical
simulations and of the study of a free energy function very similar
to eq.~(\ref{eq:FreeEnergy}).

The free energy used in \cite{baldassi-subdominant} differs from
the one of eq.~(\ref{eq:FreeEnergy}) because in the latter we do
not impose any constraint on the reference configurations $\tilde{x}$.
In \cite{baldassi-subdominant}, we analyzed the free energy using
the Replica Method at the level of a Replica Symmetric (RS) solution,
and found that there was a maximum value of $y$ for which the \emph{external
entropy} was non-negative. The external entropy is the logarithm of
the number of configurations $\tilde{x}^{\star}$ divided by $N$,
and should not take finite negative values: when it does, it means
that there is a problem in the RS assumption. Therefore, in that analysis,
we used the value $y^{\star}$ that led to a zero complexity. 

We repeated the RS computation for the unconstrained version, eq.~(\ref{eq:FreeEnergy}),
and found that as $\alpha$ increases the results become patently
unphysical. For example, the RS solution at $y=y^{\star}$ would predict
a positive local entropy even beyond the critical value $\alpha_{c}$.
Therefore, the RS assumption needs to be abandoned and at least a
$1$-step of Replica Symmetry Breaking ($1$-RSB) must be studied.
Specifically, we assumed that RSB occurred at the level of the $\tilde{x}$
variables, while we kept the RS Ansatz for the $x$ variables, and
computed the result in the limit $y\to\infty$. This appears as a
geometrically consistent assumption, in that the clusters we are analyzing
are dense and we do not expect any internal fragmentation of their
geometrical structure. All the details of the calculation are in the
Appendix~\ref{sec:APPENDIX-II-Details}. The result is a system of
$8$ coupled equations, with $\alpha$ and $S$ as control parameters
(using $S$ as control parameter instead of its conjugate $\gamma$
which was used in eq.~\ref{eq:FreeEnergy} is advantageous for the
theoretical analysis at high $\alpha$, see below).

Solving these equations still yields a negative external entropy for
all values of $\alpha$ and $S$, and studying the finite $y$ case
is numerically much more challenging in this case, to the point of
being presently unfeasible. However, the magnitude of the external
entropy is greatly reduced with respect to the analogous RS solution
at $y\to\infty$; furthermore, its value tends to zero when $S\to1$,
and all the other unphysical results of the RS solution were fixed
at this step of RSB. Additionally, the qualitative behavior of the
solution is the same as for the constrained RS version of \cite{baldassi-subdominant}.
Finally, simulation results, where available, are in remarkable agreement
with the predictions of this computation (see below, Sec.~\ref{sub:Comparison-with-theoretical}).
We therefore speculate that this solution is a reasonable approximation
to the correct solution at $y\to\infty$.

\begin{figure}

\includegraphics[width=1\textwidth]{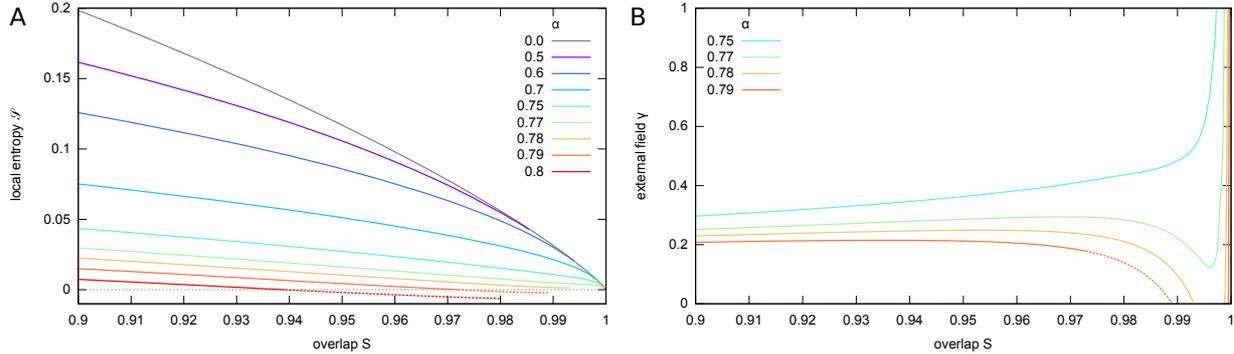}\protect\caption{\label{fig:Entropies}\textbf{A.} Local entropy vs overlap $S$, at
various values of $\alpha$. All curves tend to the $\alpha=0$ case
for sufficiently high $S$. For $\alpha\gtrsim0.77$, a gap appears,
i.e.~a region of $S$ where no solution to the saddle point equations
exists. For $\alpha\gtrsim0.79$, some parts of the curve have negative
entropy (dashed). All curves reach a plateau for sufficiently low
values of $S$ where the local entropy becomes equal to the equilibrium
entropy (not shown). \textbf{B.} Relationship between the overlap
$S$ and its conjugate parameter, the external field $\gamma$. Up
to $\alpha\lesssim0.75$, the relationship is monotonic and the convexity
does not change for all values of $S$; up to $\alpha\lesssim0.77$,
a solution exists for all $S$ but the relationship is no longer monotonic,
implying that there are regions of $S$ that can not be reached by
using $\gamma$ as an external control parameter. The gap in the solutions
that appears after $\alpha\gtrsim0.77$ is clearly signaled by the
fact that $\gamma$ reaches $0$; below $\alpha_{c}=0.83$, a second
branch of the solution always reappears at sufficiently high $S$,
starting from $\gamma=0$.}
\end{figure}

\begin{figure}

\includegraphics[width=1\textwidth]{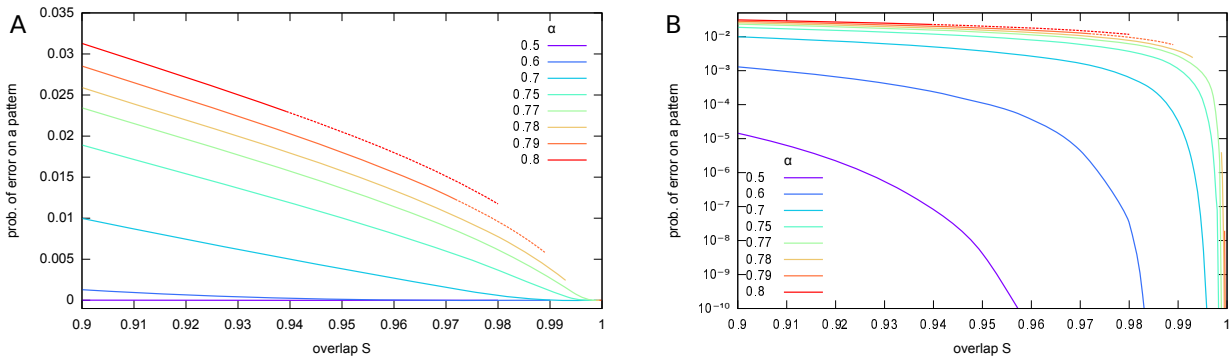}\protect\caption{\label{fig:Errors}\textbf{A.} Probability of a classification error
by the optimal reference configuration $\tilde{x}$, for various values
of $\alpha$, as a function of $S$. The dashed parts of the curves
correspond to the parts with negative local entropy (cf.~Fig.~\ref{fig:Entropies});
the curves have a gap above $\alpha\gtrsim0.77$. \textbf{B.} Same
as panel A, but in logarithmic scale on the $y$ axis, which shows
that all curves tend to zero errors for $S\to1$.}

\end{figure}

In particular, these qualitative facts hold (see Fig.~\ref{fig:Entropies}):
\begin{enumerate}
\item For all $\alpha$ below the critical value $\alpha_{c}=0.83$, the
local entropy in the region of $S\to1$ tends to the curve corresponding
to $\alpha=0$, implying that for small enough distances the region
around the ground states $\tilde{x}^{\star}$ is extremely dense (almost
all points are solutions).
\item There is a transition at $\alpha_{U}\simeq0.77$ after which the local
entropy curves are no longer monotonic; in fact, we observe the appearance
of a gap in $S$ where the system of equations has no solution. We
speculatively interpret this fact as signaling a transition between
two regimes: one for low $\alpha$ in which the ultra-dense regions
are immersed in a huge connected structure, and one at high $\alpha$
in which the structure of the sub-dominant solutions fragments into
separate regions.
\end{enumerate}
Two additional results are worth noting about the properties of the
reference configurations $\tilde{x}$ (see Appendix~\ref{sub:Reference-configurations-energy}
for the complete details):
\begin{enumerate}
\item In the limit $y\to\infty$, the the local entropy takes exactly the
same value as for the constrained case in which the $\tilde{x}$ are
required to be solutions, and the same is true for the parameters
that are common to both cases. The external entropy, however, is different.
This is true both in the RS and the $1$-RSB scenario.
\item For the unconstrained case, we can compute the probability that the
reference configuration $\tilde{x}$ makes an error on any one of
the patterns (see Fig.~\ref{fig:Errors}). It turns out that this
probability is a decreasing function of $S$ (going exponentially
to $0$ as $S\to1$) and an increasing function of $\alpha$. For
low values of $\alpha$, this probability is extremely low, such that
at finite values of $N$ the probability that $\tilde{x}$ is a solution
to the full pattern set is almost $1$.
\end{enumerate}

\subsubsection{EdMC: Local entropy as an alternative objective function in optimization\label{sub:EdMC:-Local-entropy}}

The case of the binary perceptron suggests that it is possible to
exploit the results of the sub-dominant analysis and devise a simple
and fairly general scheme for solving CSPs in a very efficient and
controlled way.

For the binary perceptron, Fig.~\ref{fig:Entropies}B shows that
up to $\alpha\lesssim0.75$ we can use $\gamma$ as a control parameter
to determine $S$, while Fig.~\ref{fig:Errors} shows that a solution
to the learning problem can be found by maximizing the local free
entropy $F\left(\tilde{x},\gamma\right)$ (eq.~(\ref{eq:local_free_entropy}))
as a function of $\tilde{x}$ at sufficiently large $\gamma$.

The reason to follow this strategy, as opposed to directly trying
to minimize the energy $H_{0}\left(\tilde{x}\right)$ (eq.~(\ref{eq:hamiltonian_generic})),
is that it turns out that the landscape of the two objective functions
is radically different: while the energy landscape can be riddled
with local minima that trap local search algorithms, the local entropy
landscape is much smoother. Therefore, a simple Monte Carlo algorithm
on the local free entropy, or ``Entropy-driven Monte Carlo'' (EdMC)
for short, is able to effectively find solutions that are very hard
to find for energy-based simulated annealing.

Furthermore, the behavior of this algorithm can --- at least in principle
--- be described in the typical case with the tools of Statistical
Mechanics, to the contrary of what is currently possible for the other
efficient solvers, which all, to some extent, resort to heuristics
(e.g.~decimation, soft decimation a.k.a.~reinforcement, etc.).

Indeed, the main practical difficulty in implementing the EdMC algorithm
sketched above is estimating the local free entropy $F\left(\tilde{x},\gamma\right)$.
We use the Belief Propagation (BP) algorithm for this, a \emph{cavity
method} algorithm that computes the answer in the context of the Bethe
Approximation. The BP algorithm is briefly explained in the Appendix~\ref{sec:APPENDIX-I-BP},
where we also reported the specific BP equations used for the particular
CSPs that we studied in this paper.

More specifically: $\tilde{x}$ is initialized at random; at each
step $F\left(\tilde{x},\gamma\right)$ is computed by the BP algorithm;
random local updates (spin flips) of $\tilde{x}$ are accepted or
rejected using a standard Metropolis rule at fixed temperature $y^{-1}$.
In practice, we found that in many regimes it suffices to use the
simple greedy strategy of a zero temperature Monte Carlo ($y=\infty$).
From a practical standpoint, it seems more important instead to start
from a relatively low $\gamma$ and increase it gradually, as one
would do in a classical annealing procedure. We call such procedure
`scoping', as it progressively narrows the focus of the local entropy
computation to smaller and smaller regions. The reason to adopt such
a strategy is easily understood by looking at the theoretical error
probability curves in Fig.~\ref{fig:Errors}.

\section{EdMC results\label{sec:EdMC-results}}

In what follows, we will describe EdMC in more detail for the two
prototypical examples introduced in Sec.~\textbf{\ref{sub:Two-prototypical-CSPs}}.

\subsection{Perceptron}

\subsubsection{Comparison with theoretical results\label{sub:Comparison-with-theoretical}}

\begin{figure}

\includegraphics[width=1\textwidth]{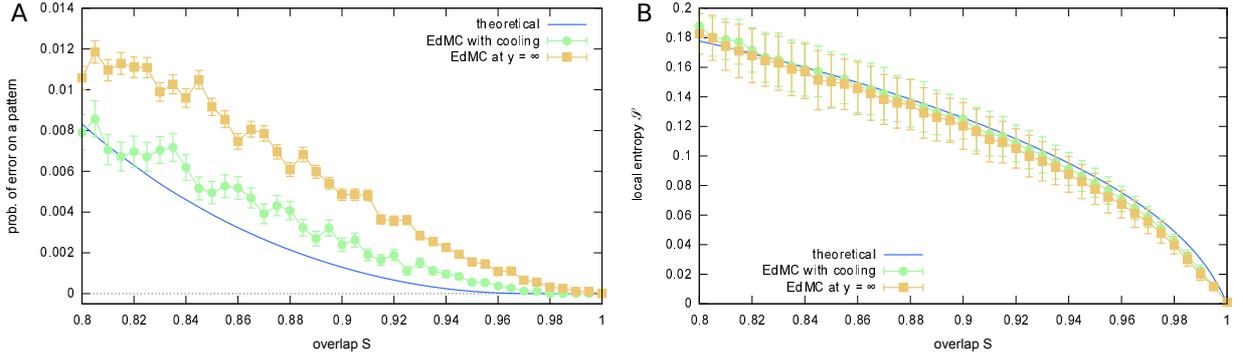}\protect\caption{\emph{\label{fig:EdMC-vs-theoretical}EdMC vs theoretical results.}
\textbf{A.} Probability of error on a pattern (cf.~Fig.~\ref{fig:Errors})
\textbf{B.} Local entropy (cf.~Fig.~\ref{fig:Entropies}A). See
text for details on the procedure. For the version with cooling, $700$
pattern sets were tested for each value of $\gamma$. For the $y=\infty$
version, $2000$ samples were used. Error bars represent standard
deviation estimates of the mean values. }

\end{figure}

We tested the theoretical results of Sec.~\ref{sub:1RSB} by running
EdMC on many samples at $N=201$ and $\alpha=0.6$, at various values
of $\gamma$ (we used $\gamma=\tanh^{-1}\left(p\right)$, varying
$p\in[0.4,0.9]$ in steps of $0.1$). In this case, since we sought
the optimum value of the free local entropy $F\left(\tilde{x},\gamma\right)$
at each $\gamma$, we did not stop the algorithm when a solution was
found. We ran the algorithm both directly at $y=\infty$ and using
a cooling procedure, in which $y$ was initialized at $5$ and increased
by a factor of $1.01$ for every $10$ accepted moves. The search
was stopped after $5N$ consecutive rejected moves. For each sample
and each polarization level, we recorded the value of the overlap
$S$, of the local entropy $\mathscr{S}$ (see eq.~\ref{eq:local_entropy}
above and eqs.~(\ref{eq:bp_overlap}) and (\ref{eq:bp_local_entropy})
in the Appendix~\ref{sec:APPENDIX-I-BP}) and of the error probability
per pattern. Then, we binned the results over the values of $S$,
using bins of width $0.005$, and averaged the results of the local
entropy and the error rate in each bin. Fig.~\ref{fig:EdMC-vs-theoretical}
shows that both values are in reasonable agreement with the theoretical
curve: the qualitative behavior is the same (in particular: the error
rate goes to zero at $S\to1$ and the entropy is positive until $S=1$,
confirming the existence of dense clusters) and the more accurate
version is closer to the theoretical values. The remaining discrepancy
could be ascribed to several factors: 1) finite size effects, since
$N$ is rather small 2) inaccuracy of the Monte Carlo sampling, which
would be fixed by lowering the cooling rate 3) inaccuracy of the theoretical
curve due to RSB effects, since we know that our solution is only
an approximation.

Note that, with these settings, the average number of errors per pattern
set is almost always less than $2$ for all points plotted in Fig.~\ref{fig:EdMC-vs-theoretical}A.
Also note that, for all values of $S$, the mode and the median of
the error distribution is at $0$, and that the average is computed
from the tails of the distribution (which explains the noise in the
graphs). Finally note that, for all samples, points corresponding
to $0$ errors were found during the Monte Carlo procedure.

\subsubsection{Comparison with standard Simulated Annealing}

We tested our method with various problem sizes $N$ and different
values of $\alpha$, and compared its performance with a standard
MCMC. The most remarkable feature of EdMC is its ability to retrieve
a solution at zero temperature in a relative small number of steps.
We found that zero temperature MCMC (Glauber dynamics) immediately
gets trapped in local minima at zero temperature, even at small $N$.
In order to find a solution with MCMC we used a simulated annealing
(SA) approach, with initial inverse temperature $y_{0}=1$, and we
increased $y$ by a factor $f_{y}$ for every $10^{3}$ accepted moves.
The factor $f_{y}$ is a cooling rate parameter that we optimized
for each problem instance (see below). Fig.~\ref{fig:en_entr_perc}
shows a comparison between a typical trajectory of SA versus EdMC
on the very same instance (EdMC is run at $y=\infty$ with $\gamma=\tanh^{-1}\left(0.6\right)$):
at first glance, it exemplifies the typical difference in the average
number of steps required to reach a solution between SA and EdMC,
which is of $4$ or $5$ orders of magnitude for small $N$. Also
note the smoothness of the EdMC trajectory in contrast to SA: the
local entropy landscape is far smoother and ensures a rapid convergence
to the region with highest density of solutions.
\begin{figure}
\includegraphics[width=0.6\textwidth]{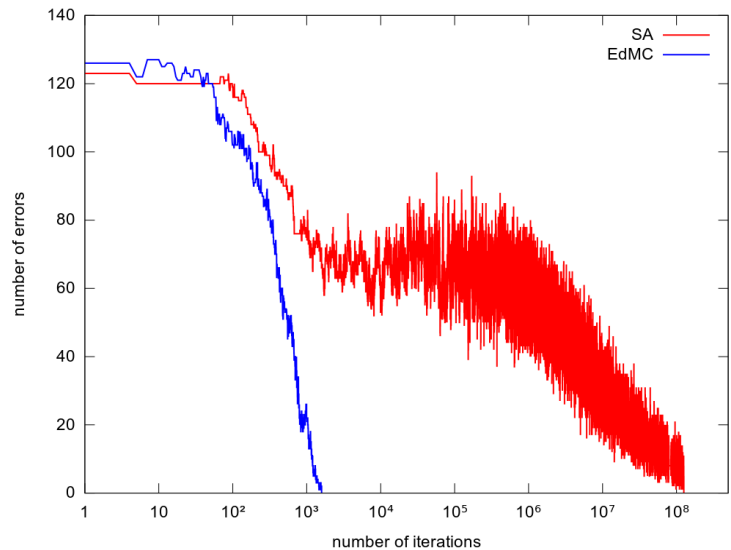}\protect\caption{\label{fig:en_entr_perc}\emph{Perceptron Learning Problem}, $N=801$,
$\alpha=0.3$. Typical trajectories of standard Simulated Annealing
on Hamiltonian (\ref{eq:hamiltonian_perc}) (red curve, right) and
Entropy-driven Monte Carlo (blue curve, left). Notice the logarithmic
scale in the $x$ axis. EdMC is run at $0$ temperature with fixed
$\gamma=\tanh^{-1}\left(0.6\right)$, SA is started at $y_{0}=1$
and run with a cooling rate of $f_{y}=1.001$ for each $10^{3}$ accepted
moves, to ensure convergence to a solution.}
\end{figure}

We studied the scaling properties of EdMC in contrast to SA, at $\alpha=0.3$
and $\alpha=0.6$, varying $N$ between $201$ and $1601$ and measuring
the number of iterations needed to reach a solution to the learning
problem.

For the SA tests, we used the following procedure: for each instance
of the problem, we tried to find a solution at some value of the cooling
rate $f_{y}$; after $10^{5}N$ consecutive rejected moves, we started
over with a reduced $f_{y}$, and repeated this until a solution was
eventually found. The values of $f_{y}$ that we used were $\left\{ 1.1\right.$,
$1.05$, $1.02$, $1.01$, $1.005$, $1.002$, $1.001$, $1.0005$,
$\left.1.0001\right\} $. This allowed us to measure the least number
of iterations required by SA to solve the problem (we only report
the number of iterations for the last attempted value of $f_{y}$).
At $\alpha=0.3$, all tested instances were solved, up to $N=1601$.
At $\alpha=0.6$, however, this procedure failed to achieve 100\%
success rate even for $N=201$ in reasonable times; at $N=401$, the
success rate was 0\% even with $f_{y}=1.0001$; at $N=1601$, using
$f_{y}=1.0001$ did not seem to yield better results than $f_{y}=1.1$.
Therefore, no data is available for SA at $\alpha=0.6$.

For the EdMC tests, we used the following procedure: we started the
algorithm at $\gamma=\tanh^{-1}\left(p\right)$ with $p=0.4$, and
ran the Monte Carlo procedure directly at $y=\infty$; if a solution
was not found, we increased $p$ by a step of $0.1$ and continued
from the last accepted configuration, up to $p=0.9$ if needed. Since
we worked at $y=\infty$, we only needed to test each spin flip at
most once before abandoning the search and increasing $p$. This procedure
allowed us to find a solution in all cases.

\begin{figure}
\includegraphics[width=1\columnwidth]{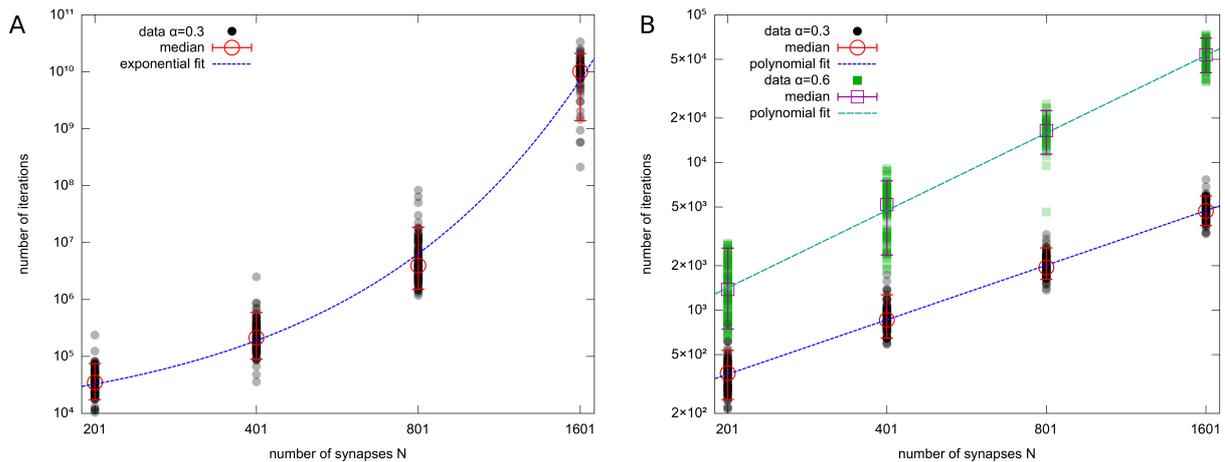}\protect\caption{\label{fig:scaling_en_perc}\emph{Perceptron Learning Problem}. Number
of iterations required to reach $0$ energy in log-log scale, as a
function of the problem size $N$. \textbf{A:} Simulated Annealing
at $\alpha=0.3$, \textbf{B:} EdMC at $\alpha=0.3$ (bottom) and $\alpha=0.6$
(top). See text for the details of the procedure. Notice the difference
in the $y$ axes scales. For both methods, $100$ samples were tested
for each value of $N$. Color shades reflect data density. Empty circles
and squares represent medians, error bars span the $5$-th to $95$-th
percentile interval. The dashed lines are fitted curves: the SA points
are fitted by an exponential curve $\exp\left(a+bN\right)$ with $a=8.63\pm0.06$,
$b=\left(8.79\pm0.08\right)\cdot10^{-3}$; the EdMC points are fitted
by two polynomial curves $aN^{b}$ with $a=0.54\pm0.04$, $b=1.23\pm0.01$
for $\alpha=0.3$, and with $a=0.14\pm0.02$, $b=1.74\pm0.02$ for
$\alpha=0.6$.}
\end{figure}

The results are shown in Fig.~\ref{fig:scaling_en_perc}, in log-log
scale. For SA (panel A in the figure), the behavior is clearly exponential
at $\alpha=0.3$, and missing altogether for $\alpha=0.6$. For EdMC
(panel B in the figure), the data is well fit by polynomial curves,
giving a scaling $\sim N^{1.23}$ for $\alpha=0.3$ and $\sim N^{1.74}$
for $\alpha=0.6$. Also note the difference of several orders of magnitude
in the ranges of the $y$ axes in the two panels. 

From the theoretical analysis, and the results shown in Figs.~\ref{fig:Entropies}
and~\ref{fig:EdMC-vs-theoretical}, it could be expected that EdMC
should be able to find a solution at least until $\alpha\sim0.75$
when the entropy curves lose their monotonicity, and therefore be
on par with reinforced Belief Propagation \cite{braunstein-zecchina}
and reinforced Max-Sum \cite{baldassi2015max} in terms of algorithmic
capacity (though being much slower in terms of absolute solving times),
if following a cooling procedure on $y$. Assessing this in detail
however would require even more extensive testing and goes beyond
the scope of the present work.

Finally, we also tested SA using a simple variant of the energy function,
in which a rectified linear function $f\left(x\right)=\max\left(0,x\right)$
is used instead of the step function $\Theta\left(x\right)$ in eq.~(\ref{eq:hamiltonian_perc}),
and verified that, while the performance indeed improves, the qualitative
picture remains unchanged.

\subsection{Extensions: the $K$-SAT case and heuristic improvements\label{sub:Extensions}}

The results we have displayed for the Binary Perceptron rely on a
general scheme that in principle can be applied to other CSPs (or
optimization problems). The main bottleneck in implementing such extensions
resides in the possibility of computing the local entropy efficiently,
e.g. by BP or some other sampling technique. As proof of concept we
apply EdMC to the very well studied case of random $K$-SAT \cite{monasson1999determining,mezard2002random,mezard2002analytic}
focusing on the non trivial case $K=4$, at various $N$ and $\alpha$.
Random $4$-SAT is characterized by three different regimes\cite{montanari2008clusters,krzakala-csp}:
For $\alpha<\alpha_{d}=9.38$ the phase is RS and the solution space
is dominated by a connected cluster of solutions with vanishing correlations
among far apart variables. For $\alpha_{d}<\alpha<\alpha_{c}=9.547$
the dominant part of the solution space brakes into an exponential
number of clusters that have an extensive internal entropy. Long range
correlations do not vanish. For $\alpha_{c}<\alpha<\alpha_{s}=9.931$
the solution space is dominated by a sub-exponential number of clusters.
Eventually for $\alpha>\alpha_{s}$ the problem becomes unsatisfiable.
The hard region for random 4-SAT is $\alpha\in[\alpha_{d,}\alpha_{s}]$,
i.e.~where long range correlations do not vanish. In such region
SA are expected to get stuck in glassy states and most of the heuristics
are known to fail.

\begin{figure}
\includegraphics[width=1\textwidth]{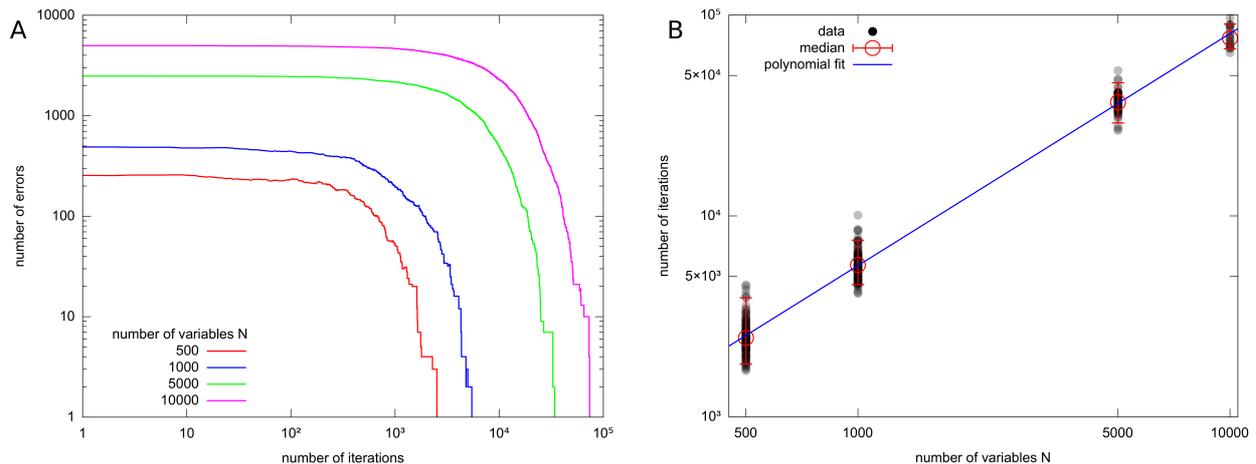}\protect\caption{\label{fig:scaling_4sat}\emph{Random $4$-SAT}, $\alpha=8.0$ (easy
phase), EdMC results with $y=\infty$ and $\gamma=\tanh^{-1}\left(0.3\right)$.
\textbf{A.} Typical trajectories at different values of $N$, from
$500$ to $10000$ (bottom to top), in log-log scale. \textbf{B.}
Number of iterations to reach $0$ energy, in log-log scale, as a
function of the problem size $N$. Color shades reflect data density.
Empty circles represent medians, error bars span the $5$-th to $95$-th
percentile interval. The data is well fitted by a polynomial curve
$aN^{b}$ (blue line), with $a=1.94\pm0.15$, $b=1.15\pm0.01$.}
\end{figure}

In the RS regime, EdMC succeeds in finding a solution in a small number
of steps, confirming the smooth nature of the objective function.
Typical trajectories for different values of $N$ are depicted in
Figure~\ref{fig:scaling_4sat}A. Figure~\ref{fig:scaling_4sat}B
shows the scaling behavior with $N$, which is polynomial ($\sim N^{1.15}$)
as in the case of the Perceptron.

In the hard phase the method suffers from the lack of convergence
of BP. Even if BP (technically the $1$-RSB solution with $m=1$)
would converge up to the condensation point $\alpha_{c}$, the addition
of the external fields prevent BP from converging even below such
point. These are expected results that could be solved by resorting
to a $1$-RSB cavity algorithm to compute the local entropy. While
certainly interesting, this is beyond the scope of this paper. For
the sake of simplicity we decided to adopt simple heuristic solutions
just to show that the overall method is effective.

\begin{figure}
\includegraphics[width=0.6\columnwidth]{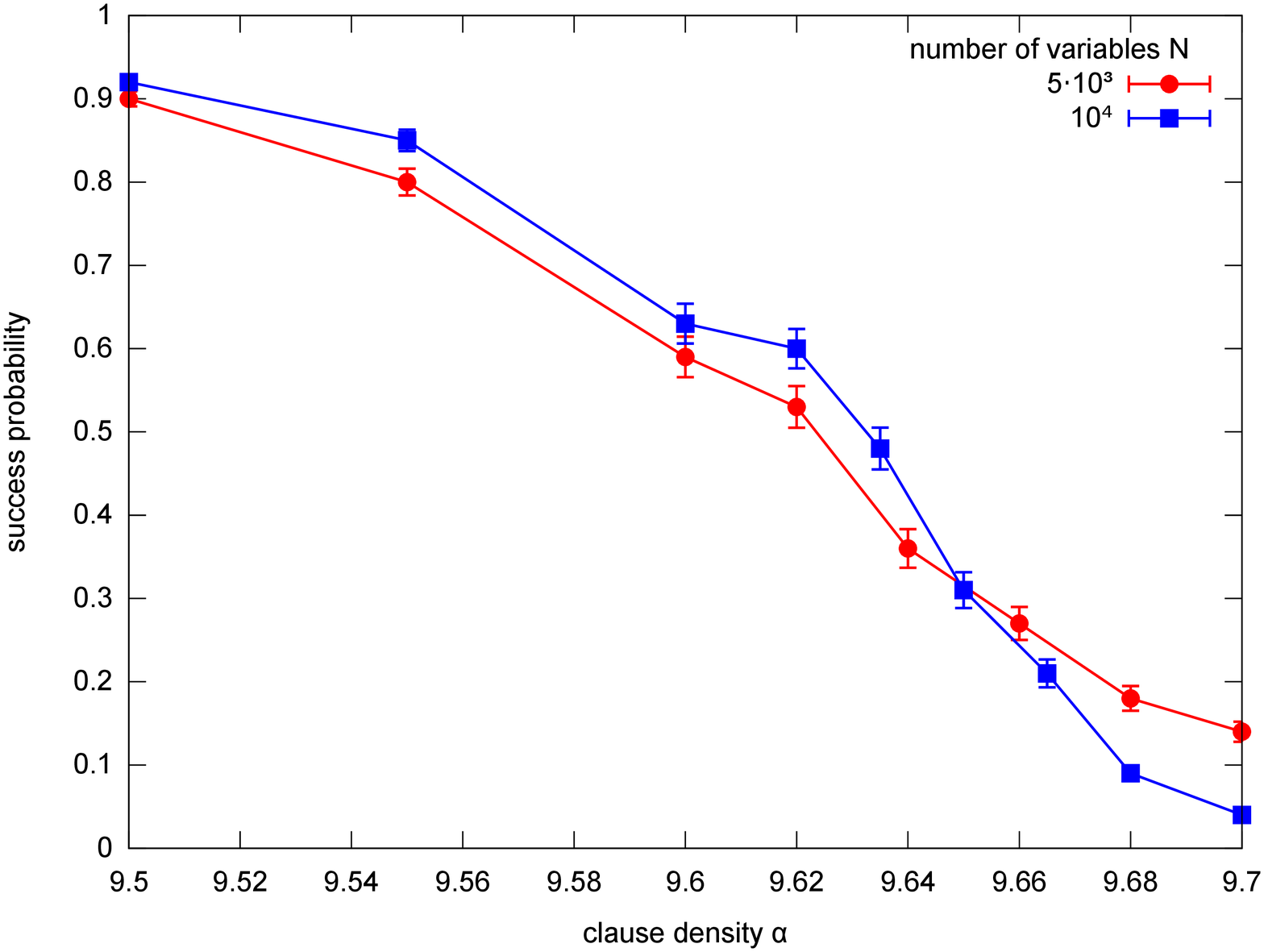}

\protect\caption{\label{fig:prob-ksat}\emph{Random 4-SAT hard phase. }Probability
of finding a solution for the faster (heuristic) EdMC algorithm on
random instances of the 4-SAT problem as a function of the clause
density $\alpha$. Data points are obtained using Algorithm \ref{alg:heuristic_algo}
(but without resorting to standard EdMC as $\left|V\right|=0$ ---
see line of code 17; rather, the algorithm is stopped and considered
to have failed in this case) and averaging over $100$ samples for
each value of $\alpha$ and each problem size. For simplicity, the
parameters of the algorithm are fixed once for all simulations, even
though they could be fine-tuned to achieve better performance: scoping
coefficient $f_{\gamma}=1.05,$ annealing coefficient $f_{y}=1.1$
and starting values $\gamma_{0}=0.1,\ y_{0}=10^{-2}$ .\textbf{ }}
\end{figure}

In cases in which BP does not converge, we take as a proxy for $F$
its average $\bar{F}$ over a sufficient number of BP iterations.
While this trick typically leads to a solution of the problem, it
has the drawback of making the overall Monte Carlo procedure slow.
To overcome this difficulty, we simply improve the choice of the flipping
step by using the information contained in the cavity marginals in
order to effectively guide the system into a state of high local density
of solutions. The heuristic turns out to be much faster and capable
of solving hard instances up to values of $\alpha$ close to the SAT/UNSAT
threshold (see Fig.~\ref{fig:prob-ksat}). The same heuristic has
also been tested in the Perceptron learning problem with excellent
results at high values of $\alpha$.

The main step of the heuristic method consists in performing an extensive
number of flipping steps at each iteration in the direction of maximum
free energy, choosing the variables to flip from the set of $x_{i}$'s
whose cavity marginals \emph{in absence} of the external field $h_{i}$
are not in agreement with the direction $\tilde{x}_{i}$ of the field
itself (see Appendix~\ref{sec:APPENDIX-I-BP} for a precise definition
of the marginals $h_{i}$). Let us call $V$ the set of such variables,
in a ranked order with respect to the difference between the external
and the cavity fields, such that the ones with the largest difference
come first. In the spirit of MCMCs, we propose a collective flip of
all the $V$ variables and compute the new value of $F$. The collective
flip is always accepted if there is an increase in free energy, otherwise
it is accepted with probability $e^{y\Delta F}$, where $\Delta F$
is the free energy difference. When a collective flip is accepted,
a new set $V$ is computed. If, on the contrary, the flips are rejected,
a new collective flip is proposed, simply eliminating the last variable
in the ranked set $V$, and the procedure is repeated until the set
is empty. In the general case this procedure quickly leads to a solution
with zero energy. If all the collective flips in $V$ are rejected,
the procedure is terminated, and a standard EdMC is started from the
last accepted values of $\tilde{x_{i}}$.

As it turns out, most of these collective moves are accepted immediately.
The interpretation is that these moves try to maximize, at each step,
the local contributions $F_{i}$'s associated to each variable $x_{i}$
in the Bethe free energy, in presence of an external field $\gamma\tilde{x}_{i}$.

As for standard EdMC, we can additionally employ an annealing strategy,
by which $y$ is progressively increased during the iteration, and
a `scoping' strategy, i.e.~a gradual increase of the external field
$\gamma$ as well. The resulting algorithm is detailed in Algorithm~\ref{alg:heuristic_algo}.

\begin{algorithm}
\protect\caption{\label{alg:heuristic_algo}Heuristic EdMC with \emph{Annealing} and
\emph{Scoping}.}
\KwIn{problem sample; parameters $t_{\mathrm{max}}$, $t_{\mathrm{step}}$, $y$, $\gamma$, $f_y$ and $f_\gamma$}
Randomly initialize $\tilde{x}^0_i$. Alternalively, run BP with $\gamma=0$ and set $\tilde{x}^0_i=\sign(h_{i})$\;
Run BP with external fields $\gamma\tilde{x}^0_i$\;
Compute free energy $F^0$ from BP fixed point ($\bar{F}^0$ if BP does not converge)\;
$t \leftarrow 0$\;
\While{$t\le t_{\mathrm{max}}$}{
  Retrieve fields $h_{i}^{t}$ ($\bar{h}_{i}^{t}$ if BP did not converge)\;  
  \lFor{$i=1$ \KwTo $N$}
    {$\Delta_{i}\leftarrow\tilde{x}_{i}^{t}\left(\gamma\tilde{x}_{i}^{t}-h_{i}^{t}\right)$}
  Collect $V=\left\{i\mid\Delta_{i}>0\right\}$ and sort it in descending order of $\Delta_i$\;
  $accepted\leftarrow \mathrm{FALSE}$\;
  \While{$\mathrm{NOT}\ accepted$}{
    Propose a flip of the $\tilde{x}_{i}^t$ for all $i \in V$, producing $\tilde{x}^{t+1}$\;
    Run BP with new proposed external fields $\gamma\tilde{x}^{t+1}_i$\;
    Compute free energy $F^{t+1}$ from BP fixed point ($\bar{F}^{t+1}$ if BP does not converge)\;
    \KwSty{with probability} $e^{y\left(F^{t+1}-F^t\right)}$ \KwSty{do} $accepted \leftarrow \mathrm{TRUE}$\;
    \If{$\mathrm{NOT}\ accepted$}{
      Remove the last element from $V$\;
      \lIf{$\left\vert{V}\right\vert=0$}{\KwSty{exit} and run EdMC with $\tilde{x}^{t}$ as initial configuration}
    }
  }
  $t\leftarrow t+1$\;
  Compute energy $E$ of configuration $\tilde{x}^{t}$\;
  \lIf{$E=0$}{retrieve solution $\tilde{x}^*=\tilde{x}^t$ and \KwSty{exit}}
  \If{$t\equiv0\ \left(\mathrm{mod}\ t_\mathrm{step}\right)$}{
    \CommentSty{Annealing}: $y \leftarrow y \times f_y$\;
    \CommentSty{Scoping}: $\gamma \leftarrow \gamma \times f_\gamma$ (run BP and update $F^t$)\;
  }
}
\end{algorithm}

Of the two strategies, annealing and scoping, the latter seems to
be far more important in practice, and in many cases crucial to produce
a quick solution: as $\gamma$ is increased, smaller regions are progressively
observed, and this focus can eventually lead the search into a given
compact cluster of solutions, a region sufficiently dense so that
Replica Symmetry locally holds. Indeed, in the typical observed trajectory
of this algorithm in the presence of the scoping dynamics, even when
BP suffers from convergence issues in the first steps, convergence
is restored when the external fields vector $\tilde{x}$ gets close
to a region of high solution density. Both strategies, scoping and
annealing, have been used to maximize the probability of success of
the algorithm in the hard phase of the $K$-SAT problem in the simulations
showed in Fig.~\ref{fig:prob-ksat}.

\section{Discussion\label{sec:Discussion}}

We recently demonstrated the relevance of a probability measure based
on the local entropy in the theoretical analysis of sub-dominant clusters
in Constraint Satisfaction Problems \cite{baldassi-subdominant}.
In the present work, we extended on the previous analysis and introduced
EdMC, a novel method for efficiently sampling over solutions in single
instances of CSPs.

At variance with standard Monte Carlo methods, we never make use directly
of the energy information: minimization of energy naturally emerges
from the maximization of local entropy. What we propose here is thus
a radically different perspective for optimization problems, based
on the possibility of estimating the local entropy, a quantity that
can effectively guide an MCMC straight into a region with a high density
of solutions, thus providing a solver. The effectiveness of our Entropy-driven
Monte Carlo may be understood in terms of a high level of smoothing
in the local entropy landscape. This is evident if we compare a zero
temperature EdMC to an energy guided SA with low cooling rate: the
former is orders of magnitude faster in terms of attempted moves,
and does not suffer from trapping in local minima. The procedure can
even be made largely more efficient by a very simple heuristic.

In order to estimate the local entropy, we rely on BP, which itself
can often be heuristically modified to become a solver (e.g.~by introducing
decimation or reinforcement schemes). However, those heuristics are
not under control, while the scheme we propose here has a clear theoretical
interpretation. BP is constructed on a cavity approximation scheme
that strongly depends on the local factorization properties of the
CSPs as well as on the global phase space structure, which is in general
dependent on the number of constraints. The validity of cavity approximation
has to be assessed for each problem at hand. It is very intriguing
to think of simpler estimations for local entropy that could not rely
on the convergence of BP equations. On the one hand, we have shown
that even when convergence is not guaranteed at all, a simple averaging
of messages could give useful information and lead to a solution in
a small number of steps, implying that even an imprecise estimate
of the local entropy is sufficient in these cases. Besides, the kinetic
accessibility of sub-domintant clusters by simple learning protocols
suggests that one could consider the general problem of engineering
simple dynamical processes governed by the measure that we introduced
in Eqs. (\ref{eq:local_free_entropy}-\ref{eq:FreeEnergy}). This
subject is currently under investigation.
\begin{acknowledgments}
CB, CL and RZ acknowledge the European Research Council for grant
n\textdegree ~267915.
\end{acknowledgments}

\appendix

\section{Belief Propagation\label{sec:APPENDIX-I-BP}}

In this section, we briefly introduce the Belief Propagation (BP)
algorithm in general, define the quantities used by EdMC, and explicitly
write the relevant expressions for the cases of the Binary Perceptron
and the $K$-SAT problems. We will make use of the notation introduced
in Sec.~\ref{sec:Sub-dominant-clusters-analysis}.

\subsection{General BP scheme\label{sub:General-BP-scheme}}

One common method for representing an instance of a CSP is to draw
a bipartite \emph{Factor Graph,} where each \emph{factor node} stands
for a constraint $\psi_{\mu}$, each \emph{variable node} stands for
a variable $x_{i}$, and each edge connects a factor node $\mu$ with
a variable node $i$ if $i\in\partial\mu$, or equivalently $\mu\in\partial i$.
This kind of graphical model representation is very helpful for understanding
the basic dynamics of message passing methods such as BP.

Belief Propagation can be considered as a tool for an efficient marginalization
of a complicated probability distribution that can be written as a
product of local interaction terms. This is obviously the case of
the Gibbs distribution associated to the Hamiltonian (\ref{eq:hamiltonian_generic}),
which reads:
\[
p_{G}(x)=\frac{1}{Z}\prod_{\mu}e^{-\beta E_{\mu}\left(x_{\partial\mu}\right)}
\]

In full generality, BP equations are a set of coupled nonlinear equations
for the \emph{cavity messages} $\left\{ u_{i\to a},\hat{u}_{a\to i}\right\} $,
which can be viewed as messages associated to each link in the Factor
Graph. For a Hamiltonian of the form (\ref{eq:hamiltonian_generic}),
the BP equations are the following: 
\begin{eqnarray}
u_{i\to\mu}\left(x_{i}\right) & \propto & \prod_{\nu\in\partial i\setminus\mu}\hat{u}_{\nu\to i}\left(x_{i}\right)\label{eq:bp_var}\\
\hat{u}_{\mu\to i}\left(x_{i}\right) & \propto & \sum_{\left\{ x_{j}\right\} _{j\neq i}}e^{-\beta E_{\mu}\left(x_{\partial\mu}\right)}\prod_{j\in\partial\mu\setminus i}u_{j\to\mu}\left(x_{j}\right)\label{eq:bp_node}
\end{eqnarray}

A common way of solving these equations is by iteration, until convergence
to a fixed point $\left\{ u_{i\to\mu}^{\star},\hat{u}_{\mu\to i}^{\star}\right\} $
is established. Fixed point messages may then be used to compute local
joint marginals and other interesting macroscopic quantities such
as the average energy and the free energy of the system. As explained
in Sec.~\ref{sub:Extensions}, one can also extract useful information
from cavity messages themselves, and rely on them to construct a very
efficient optimization procedure for the free energy. Of course, approximations
for the true marginals and free energy can be obtained by means of
instantaneous values $\left\{ u_{i\to\mu}^{t},\hat{u}_{\mu\to i}^{t}\right\} $
at a given computing time $t$, or by averaging them over the course
of the BP iterations: we exploit this fact when dealing with regimes
where the convergence of BP is hard if not impossible (see Sec.~\ref{sub:Extensions}).

Non-cavity marginals over variables can be computed at the fixed point
as:
\begin{equation}
u_{i}\left(x_{i}\right)\propto\prod_{\nu\in\partial i}\hat{u}_{\nu\to i}\left(x_{i}\right)\label{eq:bp_marginal}
\end{equation}

The modified system obtained by adding the interaction term $\gamma\tilde{x}\cdot x$
of eq.~(\ref{eq:FreeEnergy}) introduces additional terms to eqs.~(\ref{eq:bp_var})
and (\ref{eq:bp_marginal}):
\begin{eqnarray}
u_{i\to\mu}\left(x_{i};\tilde{x}_{i}\right) & \propto & e^{\gamma\tilde{x}_{i}x_{i}}\prod_{\nu\in\partial i\setminus\mu}\hat{u}_{\nu\to i}\left(x_{i}\right)\label{eq:bp_var_ext}\\
u_{i}\left(x_{i};\tilde{x}_{i}\right) & \propto & e^{\gamma\tilde{x}_{i}x_{i}}\prod_{\nu\in\partial i}\hat{u}_{\nu\to i}\left(x_{i}\right)\label{eq:bp_marginal_ext}
\end{eqnarray}

In the heuristic version of EdMC presented in Sec.~\ref{sub:Extensions}
(Algorithm~\ref{alg:heuristic_algo}), we consider the case of binary
spins $x_{i}=\pm1$ and we use the cavity magnetization fields in
absence of the external fields. These are defined from expression~(\ref{eq:bp_marginal})
as:
\begin{equation}
h_{i}=\tanh^{-1}\left(u_{i}\left(1\right)-u_{i}\left(-1\right)\right)\label{eq:bp_cavity_mag_fields}
\end{equation}

The local free entropy $F\left(\tilde{x},\gamma\right)$ of eq.~(\ref{eq:local_free_entropy})
can be computed from the fixed point messages in the zero-temperature
limit $\beta\to\infty$ in terms of purely local contributions from
variables, edges and factor nodes:
\begin{equation}
F\left(\tilde{x},\gamma\right)=\frac{1}{N}\sum_{\mu}\left(F_{\mu}\left(\tilde{x},\gamma\right)+\sum_{i\in\partial\mu}F_{i\to\mu}\left(\tilde{x},\gamma\right)\right)-\frac{1}{N}\sum_{i}\left(\left|\partial i\right|-1\right)F_{i}\left(\tilde{x},\gamma\right)\label{eq:bp_free_entropy}
\end{equation}

where:
\begin{eqnarray}
F_{\mu}\left(\tilde{x},\gamma\right) & = & \log\left(\sum_{\left\{ x_{\partial\mu}:E\left(x_{\partial\mu}\right)=0\right\} }\prod_{i\in\partial\mu}u_{i\to\mu}\left(x_{i};\tilde{x}_{i}\right)\right)\\
F_{i\to\mu}\left(\tilde{x},\gamma\right) & = & \log\left(\sum_{x_{i}}e^{\gamma\tilde{x}_{i}x_{i}}\prod_{\nu\in\partial i\setminus\mu}\hat{u}_{\nu\to i}\left(x_{i}\right)\right)\\
F_{i}\left(\tilde{x},\gamma\right) & = & \log\left(\sum_{x_{i}}e^{\gamma\tilde{x}_{i}x_{i}}\prod_{\mu\in\partial i}\hat{u}_{\mu\to i}\left(x_{i}\right)\right)
\end{eqnarray}

The overlap $S\left(\tilde{x},\gamma\right)=\frac{1}{N}\left\langle \tilde{x}\cdot x\right\rangle $
and the local entropy $\mathscr{S}\left(\tilde{x},\gamma\right)$
can be computed as: 
\begin{eqnarray}
S\left(\tilde{x},\gamma\right) & = & \frac{1}{N}\sum_{i}\tilde{x}_{i}\sum_{x_{i}}x_{i}u_{i}\left(x_{i};\tilde{x}_{i}\right)\label{eq:bp_overlap}\\
\mathscr{S}\left(\tilde{x},\gamma\right) & = & F\left(\tilde{x},\gamma\right)-\gamma S\left(\tilde{x},\gamma\right)\label{eq:bp_local_entropy}
\end{eqnarray}
These expressions are used in Sec.~\ref{fig:EdMC-vs-theoretical},
where $F\left(\tilde{x},\gamma\right)$ is optimized over $\tilde{x}$
and they are averaged over many realizations of the patterns to compare
them to the theoretical expression of eq.~(\ref{eq:local_entropy}).

\subsection{BP for the binary perceptron}

The BP equations for a given instance $\left(\xi^{\mu},\sigma^{\mu}\right),\;\mu=1...\alpha N$
are most easily written in terms of cavity magnetizations $m_{i\to\mu}\propto u_{i\to\mu}\left(+1\right)-u_{i\to\mu}\left(-1\right)$
and $\hat{m}_{\mu\to i}\propto\hat{u}_{\mu\to i}\left(+1\right)-\hat{u}_{\mu\to i}\left(-1\right)$.
Note that it is always possibile to set $\forall\mu:\,\sigma^{\mu}=1$
without loss of generality, by means of the simple gauge transformation
$\xi_{i}^{\mu}\to\sigma^{\mu}\xi_{i}^{\mu}$. With these simplification,
equations (\ref{eq:bp_var},\ref{eq:bp_node}) become:

\begin{eqnarray}
m_{i\to\mu} & = & \tanh\left(\sum_{\nu\neq\mu}\tanh^{-1}\left(\hat{m}_{\nu\to i}\right)\right)\label{eq:bp_perc_var}\\
\hat{m}_{\mu\to i} & = & \frac{\sum_{s=-\xi_{i}}^{N-1}D_{\mu\to i}\left(s\right)-\sum_{s=\xi_{i}}^{N-1}D_{\mu\to i}\left(s\right)}{\sum_{s=-\xi_{i}}^{N-1}D_{\mu\to i}\left(s\right)+\sum_{s=\xi_{i}}^{N-1}D_{\mu\to i}\left(s\right)}\label{eq:bp_perc_node}
\end{eqnarray}
where 
\begin{equation}
D_{\mu\to i}\left(s\right)=\sum_{\left\{ x_{j}\right\} _{j\neq i}}\delta\left(s,\sum_{j}x_{j}\xi_{j}\right)\prod_{j\neq i}\frac{\left(1+x_{j}m_{j\to\mu}\right)}{2}\label{eq:cavity_field_perc}
\end{equation}
is the convolution of the all cavity messages $m_{j\to\mu}$ impinging
on the pattern node $\mu$, except for $m_{i\to\mu}$: it is thus
the (cavity) distribution of the total synaptic input for pattern
$\mu$, in absence of the synapse $i$. The complexity of the second
update is at most $O\left(N^{2}\right)$ with an appropriate pre-computation
of cavity convolutions. When one deals with the case of $N\gg1$ and
an extensive number of patterns, a common and simple strategy is to
adopt a Gaussian approximation $\tilde{D}_{\mu\to i}\left(s\right)=\frac{1}{b_{\mu\to i}}G\left(\frac{s-a_{\mu\to i}}{b_{\mu\to i}}\right)$
for the distribution $D_{\mu\to i}\left(s\right)$, where $G\left(s\right)$
denotes the normal distribution. It suffices then to compute the mean
$a_{\mu\to i}$ and variance $b_{\mu\to i}^{2}$ of the distribution
$\tilde{D}_{\mu\to i}\left(s\right)$, whose dependence on cavity
messages $m_{j\to\mu}$ is easily determined from the central limit
theorem:
\begin{eqnarray}
a_{\mu\to i} & = & \sum_{j\neq i}\xi_{j}^{\mu}m_{j\to\mu}\\
b_{\mu\to i}^{2} & = & \sum_{j\neq i}\left(1-m_{j\to\mu}^{2}\right)
\end{eqnarray}
(analogous non-cavity quantities $a_{\mu}$ and $b_{\mu}$ are computed
by summing over all indices $j$). By doing so, equation (\ref{eq:bp_perc_node})
becomes: 
\begin{eqnarray}
\hat{m}_{\mu\to i} & = & \xi_{i}\,g\left(a_{\mu\to i},b_{\mu\to i}\right)\label{eq:bp_perc_node_gauss}
\end{eqnarray}
where 
\begin{equation}
g\left(a,b\right)=\frac{H\left(\frac{a-1}{b}\right)-H\left(\frac{a+1}{b}\right)}{H\left(\frac{a-1}{b}\right)+H\left(\frac{a+1}{b}\right)}
\end{equation}
and we used the function $H\left(x\right)=\frac{1}{2}\mathrm{erfc}\left(\frac{x}{\sqrt{2}}\right)$.

The free-entropy $F_{\mathrm{perc}}$ can be easily obtained by eq.~(\ref{eq:bp_free_entropy})
putting $\gamma=0$. In our Gaussian approximation, the expression
can be written as:
\begin{eqnarray}
F_{\mathrm{perc}} & = & \frac{1}{N}\sum_{\mu}\log\left(H\left(\frac{a_{\mu}}{b_{\mu}}\right)\right)-\frac{1}{N}\sum_{i,\mu}\log\left(1+m_{i\to\mu}\hat{m}_{\mu\to i}\right)\nonumber \\
 &  & +\frac{1}{N}\sum_{i}\log\left[\prod_{\mu}\left(1+\hat{m}_{\mu\to i}\right)+\prod_{\mu}\left(1-\hat{m}_{\mu\to i}\right)\right]\label{eq:free_entropy_perc}
\end{eqnarray}

The total number of solutions for a given instance can be determined
by means of the entropy: 
\begin{eqnarray}
\mathscr{S}_{\mathrm{perc}} & = & \frac{1}{N}\sum_{\mu}\log\left(H\left(\frac{a_{\mu}}{b_{\mu}}\right)\right)-\frac{1}{N}\sum_{i,\mu}\left[\frac{1+m_{i}}{2}\log\left(\frac{1+m_{i\to\mu}}{2}\right)+\frac{1-m_{i}}{2}\log\left(\frac{1-m_{i\to\mu}}{2}\right)\right]\nonumber \\
 &  & +\frac{M-1}{N}\sum_{i}\left[\frac{1+m_{i}}{2}\log\left(\frac{1+m_{i}}{2}\right)+\frac{1-m_{i}}{2}\log\left(\frac{1-m_{i}}{2}\right)\right]\label{eq:entropy_perc}
\end{eqnarray}

Consistently with the theoretical predictions, BP equations always
converge for $\alpha<\alpha_{c}$. Indeed, Replica Symmetry holds
with the identification of states as the dominant isolated configurations,
this being evident in the proportionality relation between RS and
RSB free-energy \cite{krauth-mezard}, with an intra-state overlap
(RSB parameter) $q_{1}$ equal to $1$. The entropy decreases monotonically
with $\alpha$ and, provided $N$ is high enough, it vanishes at the
critical threshold $\alpha_{c}\sim0.833$ \cite{krauth-mezard}.

Very interestingly, if one slightly modifies the original equations
with the introduction of a `reinforcement term', much similar to a
sort of smooth decimation procedure, BP becomes a very efficient solver
that is able to find a solution with probability $1$ up to $\alpha\sim0.74$
\cite{braunstein-zecchina}. Reinforced BP equations can be further
simplified, this leading to a dramatic reduction in update complexity.
The resulting on-line algorithms, SBPI \cite{baldassi-et-all-pnas}
and CP+R \cite{baldassi-2009} are able to achieve a slightly lower
capacity ($\alpha\sim0.69$). As we pointed out in the Introduction,
Sec.~\ref{sec:Introduction}, the performances of all the cited algorithms
seem to be directly affected by large sub-dominant clusters: these
high entropy states happen to be easily accessible, while the dominant
isolated solutions are very hard (if not impossible) to find efficiently
by simple learning methods.

Introducing the external fields $\gamma\tilde{x_{i}}$ of eq.~(\ref{eq:local_num_sol_soft})
is very simple (cf.~eqs.~(\ref{eq:bp_var_ext}) and (\ref{eq:bp_marginal_ext})).
Eq.~(\ref{eq:bp_perc_var}) for the cavity magnetization is modified
as:
\begin{equation}
m_{i\to\mu}=\tanh\left(\sum_{\nu\neq\mu}\tanh^{-1}\left(\hat{m}_{\nu\to i}\right)+\gamma\tilde{x_{i}}\right)\label{eq:bp_perc_var_fields}
\end{equation}
and similarly for the total magnetization: $m_{i}=\tanh\left(\sum_{\mu}\tanh^{-1}\left(\hat{m}_{\mu\to i}\right)+\gamma\tilde{x_{i}}\right)$.
The local free entropy is also simply given by the expression $F_{\mathrm{perc}}\left(\tilde{x},\gamma\right)=\mathscr{S}_{\mathrm{perc}}\left(\tilde{x},\gamma\right)+\gamma S\left(\tilde{x},\gamma\right)$
using eqs.~(\ref{eq:entropy_perc}) and (\ref{eq:bp_overlap}) with
the modified magnetizations. The cavity magnetization fields in absence
of the external fields, eq.~(\ref{eq:bp_cavity_mag_fields}), are:
\begin{equation}
h_{i}=\sum_{\mu}\tanh^{-1}\left(\hat{m}_{\mu\to i}\right)
\end{equation}

\subsection{BP for $K$-SAT}

The Belief Propagation equations for the $K$-SAT problem are most
easily written with a parametrization of the BP messages$\left\{ u_{i\to\mu},\hat{u}_{\mu\to i}\right\} $
in terms of the quantities $\left\{ \zeta_{i\to\mu},\eta_{\mu\to i}\right\} $,
where $\zeta_{i\to\mu}$ is the probability that $x_{i}$ does not
satisfy clause $\mu$ in absence of this clause, and $\eta_{\mu\to i}$
is the probability that all the variables in clause $\mu$ except
$i$ violate the clause. With this choice, and calling $V(\mu)$ the
set of variables in the constraint $\mu$, equation (\ref{eq:bp_node})
simply becomes:
\begin{equation}
\eta_{\mu\to i}=\prod_{j\in V\left(\mu\right)\setminus i}\zeta_{j\to\mu}\label{eq:bp_eta_sat}
\end{equation}

Equation (\ref{eq:bp_var}) for the variable node update is slightly
more involved:
\begin{equation}
\zeta_{i\to\mu}=\frac{\prod_{\nu\in V_{\mu}^{s}\left(i\right)}\left(1-\eta_{\nu\to i}\right)}{\prod_{\nu\in V_{\mu}^{s}\left(i\right)}\left(1-\eta_{\nu\to i}\right)+\prod_{\nu\in V_{\mu}^{u}\left(i\right)}\left(1-\eta_{\mu\to i}\right)}\label{eq:bp_zeta_sat}
\end{equation}
where $V_{\mu}^{s}\left(i\right)$ (resp.~$V_{\mu}^{u}\left(i\right)$)
is the set of clauses $\nu$ in which variable $i$ is involved with
a coupling $J_{i}^{\nu}\ne J_{i}^{\mu}$ (resp.~$J_{i}^{\nu}=J_{i}^{\mu}$).

As for the perceptron, the free-entropy is obtained from expression
(\ref{eq:bp_free_entropy}) at $\gamma=0$: 
\begin{eqnarray}
F_{\mathrm{SAT}} & = & \frac{1}{N}\sum_{\mu}\log\left(1-\prod_{i\in V(\mu)}\zeta_{i\to\mu}\right)-\frac{1}{N}\sum_{i}\sum_{\mu\in\partial i}\log\left(1-\zeta_{i\to\mu}^{2}\right)+\nonumber \\
 &  & +\frac{1}{N}\sum_{i}\log\left[\prod_{\nu\in V^{+}\left(i\right)}\left(1-\eta_{\nu\to i}\right)+\prod_{\nu\in V^{-}\left(i\right)}\left(1-\eta_{\nu\to i}\right)\right]\label{eq:free_entropy_sat}
\end{eqnarray}
where $V^{+}\left(i\right)$ (resp.~$V^{-}\left(i\right)$) is the
set of all clauses $\mu$ in which variable $i$ is involved with
$J_{i}^{\mu}=-1$ (resp.~$J_{i}^{\mu}=1$). Analogously, the entropy
is obtained from the following expression, which only depends upon
the messages $\eta_{\mu\to i}$:

\begin{eqnarray}
\mathscr{S}_{\mathrm{SAT}} & = & \frac{1}{N}\sum_{\mu}\log\left[\prod_{i\in V\left(\mu\right)}\left(\prod_{\nu\in V_{\mu}^{s}\left(i\right)}\left(1-\eta_{\nu\to i}\right)+\prod_{\nu\in V_{\mu}^{u}\left(i\right)}\left(1-\eta_{\nu\to i}\right)\right)-\prod_{i\in V\left(\mu\right)}\left(\prod_{\nu\in V_{\mu}^{u}\left(i\right)}\left(1-\eta_{\nu\to i}\right)\right)\right]\nonumber \\
 &  & +\frac{1}{N}\sum_{i}\left(1-\left|\partial i\right|\right)\log\left[\prod_{\nu\in V^{+}\left(i\right)}\left(1-\eta_{\nu\to i}\right)+\prod_{\nu\in V^{-}\left(i\right)}\left(1-\eta_{\nu\to i}\right)\right]\label{eq:entropy_sat}
\end{eqnarray}

In the chosen parametrization, the external fields $\gamma\tilde{x}_{i}$
may be easily introduced by means of $N$ additional single-variable
`soft clauses' $\tilde{C}_{i}$, which send fixed cavity messages
to their respective variables, taking the value: 
\begin{equation}
\eta_{\tilde{C}_{i}\to i}=\frac{2\tanh\gamma}{1+\tanh\gamma}\label{eq:ext_field_sat}
\end{equation}
with the restriction that $\tilde{C}_{i}\in V^{+}\left(i\right)$
(resp.~$\tilde{C}_{i}\in V^{-}\left(i\right)$) if $\tilde{x}_{i}=+1$
(resp.~$\tilde{x}_{i}=-1$).

If we call $\tilde{V}^{+}\left(i\right)$, $\tilde{V}^{-}\left(i\right)$
the new set of clauses, enlarged so as to contain the $\tilde{C}_{i}$,
the total magnetization is given by: 
\begin{equation}
m_{i}=\frac{\prod_{\nu\in\tilde{V}^{-}\left(i\right)}\left(1-\eta_{\nu\to i}\right)-\prod_{\nu\in\tilde{V}^{+}\left(i\right)}\left(1-\eta_{\nu\to i}\right)}{\prod_{\nu\in\tilde{V}^{-}\left(i\right)}\left(1-\eta_{\nu\to i}\right)+\prod_{\nu\in\tilde{V}^{+}\left(i\right)}\left(1-\eta_{\mu\to i}\right)}\label{eq:mag_tot_sat}
\end{equation}
The cavity magnetization fields in absence of the external fields,
eq.~(\ref{eq:bp_cavity_mag_fields}), are simply given by:
\begin{equation}
h_{i}=\tanh^{-1}\left(m_{i}\right)-\gamma\tilde{x}_{i}\label{eq:mag_cav_sat}
\end{equation}

Convergence properties of equations (\ref{eq:bp_eta_sat},\ref{eq:bp_zeta_sat})
on single instances are deeply related to the Replica Symmetry Breaking
scenario briefly discussed in the preceding section. The onset of
long range correlations in clustered RSB phase prevents BP from converging.

While RSB limits the usefulness of BP (as well as reinforced BP) at
high $\alpha$, ideas from the $1$-RSB cavity method, when applied
to the single case without the averaging process, have led to the
introduction of a powerful heuristic algorithm, Survey Propagation
(SP) \cite{mezard2002analytic,braunstein2005survey}, which is able
to solve $K$-SAT instances in the hard phase, almost up to the UNSAT
threshold.

\section{Details of the large deviations analysis for the binary Perceptron
Learning problem\label{sec:APPENDIX-II-Details}}

In this section, we provide all technical details of the analysis
of the reweighted free energy function of eq.~(\ref{eq:FreeEnergy})
of Sec.~\ref{sec:Sub-dominant-clusters-analysis} for the case of
the Perceptron Learning problem with binary synapses of Sec.~\ref{par:Binary-perceptron}.
The results of this analysis are presented in Sec.~\ref{sub:Comparison-with-theoretical}.

However, the notation in this section will differ at times from the
one in the main text, to make it more similar to the one used in~\cite{baldassi-subdominant},
and so that this section is mostly self-contained.

\subsection{Setting the problem and the notation\label{sec:APPENDIX-II-Intro}}

We generate patterns by drawing the inputs as random i.i.d. variables
$\xi_{i}\in\left\{ -1,+1\right\} $ with distribution $P\left(\xi_{i}\right)=\frac{1}{2}\delta\left(\xi_{i}-1\right)+\frac{1}{2}\delta\left(\xi_{i}+1\right)$.
Without loss of generality, we assume that the desired output of all
patterns is $1$.

We consider the learning problem of correctly classifying a set of
$\alpha N$ patterns $\left\{ \xi^{\mu}\right\} $ (where $\mu=1,\dots,\alpha N$).
We define, for any vector of synaptic weights $W=\left\{ W_{i}\right\} _{i=1,\dots,N}$
with $W_{i}\in\left\{ -1,+1\right\} $ the quantity 
\begin{equation}
\mathbb{X}_{\xi}\left(W\right)=\prod_{\mu}\Theta\left(\frac{1}{\sqrt{N}}\sum_{i}W_{i}\xi_{i}^{\mu}\right)
\end{equation}
(where we used $\Theta$ to represent the Heaviside step function:
$\Theta\left(x\right)=1$ if $x\ge0$, $0$ otherwise) such that the
solutions of the learning problem are described by $\mathbb{X}_{\xi}\left(W\right)=1$.
The factor $1/\sqrt{N}$ has been added to account for the scaling
of the effective fields. 

Note that throughout this section we will use the index $i$ for the
synapses and $\mu$ for the patterns. In all sums and products, the
summation ranges are assumed implicitly, e.g.~$\sum_{i}\equiv\sum_{i=1}^{N}$.
Also, all integrals are assumed to be taken over $\mathbb{R}$ unless
otherwise specified. The letters $a$, $b$, $c$ and $d$ will be
reserved for replica indices (see below). Finally, when we will make
the $1$-RSB Ansatz, we will also use $\alpha$, $\beta$, $\alpha^{\prime}$
and $\beta^{\prime}$ for the replica indices; it should be clear
from the context that these do not refer to the capacity $\alpha$
or the inverse temperature $\beta$ in those cases.

We write the number of solutions as:
\begin{equation}
\mathcal{N}_{\xi}=\sum_{\left\{ W\right\} }\mathbb{X}_{\xi}\left(W\right)
\end{equation}

We want to study the solution space of the problem in the following
setting: we consider a set of reference configurations, each of which
has an associated set of solutions that are constrained to have a
certain overlap $S$ with it.

In the following, we indicate with $\tilde{W}$ the reference configurations,
and with $W$ the other solutions. In general we use a tilde for all
the quantities that refer to the reference configurations.

Let us define then:

\begin{equation}
\mathcal{N}_{\xi}\left(\tilde{W},S\right)=\sum_{\left\{ W\right\} }\mathbb{X}_{\xi}\left(W\right)\delta\left(\sum_{i}W_{i}\tilde{W}_{i}-SN\right)\label{eq:local_num_sol_hard}
\end{equation}
(where $\delta$ is the Dirac-delta distribution\footnote{we should have used a Kronecker delta symbol here, but this abuse
of notation comes in handy since in the following we will use integrals
instead of sums for the weights}), i.e.~the number of solutions $W$ that have overlap $S$ with
(or equivalently, distance $\frac{1-S}{2}$ from) a reference configuration
$\tilde{W}$. We then introduce the following quenched free energy:
\begin{equation}
\mathscr{F}\left(S,y\right)=-\frac{1}{Ny}\left\langle \log\left(\Omega\left(S,y\right)\right)\right\rangle _{\left\{ \xi^{\mu}\right\} }=-\frac{1}{Ny}\left\langle \log\left(\sum_{\left\{ \tilde{W}\right\} }\mathcal{N}_{\xi}\left(\tilde{W},S\right)^{y}\right)\right\rangle _{\left\{ \xi^{\mu}\right\} }\label{eq:F_def}
\end{equation}
where $\Omega\left(S,y\right)$ is the partition function and $y$
has the role of an inverse temperature. This is the free energy density
of a system where the configuration is described by $\tilde{W}$ and
the energy is given by minus the entropy of the other solutions with
overlap $S$ from it.

This expression is almost equivalent to eq.~\ref{eq:FreeEnergy}
in the main text, except that we used the overlap $S$ as a control
parameter instead of the coupling $\gamma$, by using a hard constraint
(the delta distribution in eq.~\ref{eq:local_num_sol_hard}) rather
than a soft constraint (the exponential in eq.~\ref{eq:local_num_sol_soft}).
The two parameters are conjugates. The main advantage of using $\gamma$
is that it is easier to implement using the BP algorithm, which is
why we used it for the EdMC algorithm. The advantage of using $S$
is that it provides a more general description: while in large portions
of the phase space the relationship between $\gamma$ and $S$ is
bijective (and thus the two systems are equivalent), some regions
of the phase space at large $\alpha$ can only be fully explored with
by constraining $S$, and thus we have used this system for the theoretical
analysis.

The main goal is that of studying the ground states of the system,
i.e.~taking the limit of $y\to\infty$. This limit allows us to seek
the reference configuration $\tilde{W}$ for which the number of solutions
at overlap $S$ with it is maximal, and to derive an expression for
the entropy $\mathscr{S}\left(S,y\right)=\left\langle \log\mathcal{N}_{\xi}\left(\tilde{W},S\right)\right\rangle $
of the surrounding solutions, which we call local entropy. As we shall
see, we find that $\mathscr{S}\left(S,\infty\right)$ is always positive
for $\alpha<\alpha_{c}$ when $S\to1$, indicating the presence of
dense clusters of solutions.

In the remainder, we will generally use the customary notation $\int d\mu\left(W\right)$
or $\int\prod_{i}d\mu\left(W_{i}\right)$ (in stead of $\sum_{W}$
or $\sum_{\left\{ W\right\} }$) to denote the integral over possible
values of the weights; since we assume binary weights, we will have:
\[
d\mu\left(W\right)=\left(\delta\left(W-1\right)+\delta\left(W+1\right)\right)dW
\]

\subsection{Entropy and Complexity}

\subsubsection{Replica trick\label{sub:Replica-trick}}

In order to compute the quantity of eq.~(\ref{eq:F_def}), we use
the replica trick. We will use $n$ to denote the number of replicas
of the reference configurations, and the letters $c$ and $d$, with
$c,d\in\left\{ 1,\dots,n\right\} $, as their replica indices.

We will also write $\mathcal{N}\left(\tilde{W},S\right)^{y}$ as a
product of $y$ ``local'' replicas. Note that we will have a different
set of local replicas for each replicated reference configuration,
so that the local replicas are $yn$ in total. We use the indices
$a$ and $b$ to denote these local replicas (each of which will also
have a reference replica index $c$), i.e.~$a,b\in\left\{ 1,\dots,y\right\} $.

Therefore, we need to compute:
\begin{eqnarray}
 &  & \lim_{n\to0}\left\langle \Omega\left(S,y\right)^{n}\right\rangle _{\left\{ \xi^{\mu}\right\} }=\label{eq:omega1}\\
 &  & =\lim_{n\to0}\left\langle \int\prod_{ic}d\mu\left(\tilde{W}_{i}^{c}\right)\int\prod_{ica}d\mu\left(W_{i}^{ca}\right)\prod_{ca}\mathbb{X}_{\xi}\left(W^{ca}\right)\prod_{ca}\delta\left(\sum_{i}W_{i}^{ca}\tilde{W}_{i}^{c}-SN\right)\right\rangle _{\left\{ \xi^{\mu}\right\} }\nonumber 
\end{eqnarray}

As a first step, we substitute the arguments of the theta functions
in the $\mathbb{X}_{\xi}$ terms via Dirac-delta functions:
\begin{equation}
\prod_{ca\mu}\Theta\left(\frac{1}{\sqrt{N}}\sum_{i}W_{i}^{ca}\xi_{i}^{\mu}\right)=\int\prod_{ca\mu}d\lambda_{\mu}^{ca}\delta\left(\lambda_{\mu}^{ca}-\frac{1}{\sqrt{N}}\sum_{i}W_{i}^{ca}\xi_{i}^{\mu}\right)\prod_{ca\mu}\Theta\left(\lambda_{\mu}^{ca}\right)\label{eq:thetas_expansion}
\end{equation}

Then, we expand the delta functions using their integral representation:
\begin{equation}
\delta\left(\lambda_{\mu}^{ca}-\frac{1}{\sqrt{N}}\sum_{i}W_{i}^{ca}\xi_{i}^{\mu}\right)=\int\frac{d\hat{\lambda}_{\mu}^{ca}}{2\pi}\exp\left(i\hat{\lambda}_{\mu}^{ca}\lambda_{\mu}^{ca}-i\hat{\lambda}_{\mu}^{ca}\frac{1}{\sqrt{N}}\sum_{i}W_{i}^{ca}\xi_{i}^{\mu}\right)\label{eq:deltas_expansion}
\end{equation}
With these, we can factorize the expression where the patterns are
involved, so we can compute the averages over the patterns independently
for each $\mu$ and for each $i$, and expand for large $N$:
\begin{eqnarray}
 &  & \prod_{i}\int\left(P\left(\xi_{i}^{\mu}\right)d\xi_{i}^{\mu}\right)\exp\left(-\frac{i}{\sqrt{N}}\xi_{i}^{\mu}\left(\sum_{ca}W_{i}^{ca}\hat{\lambda}_{\mu}^{ca}\right)\right)=\nonumber \\
 &  & \simeq\exp\left(-\frac{1}{2N}\sum_{i}\left(\sum_{ca}W_{i}^{ca}\hat{\lambda}_{\mu}^{ca}\right)^{2}\right)\nonumber \\
 &  & =\exp\left(-\frac{1}{2}\left(\sum_{cadb}\hat{\lambda}_{\mu}^{ca}\hat{\lambda}_{\mu}^{db}\left(\frac{1}{N}\sum_{i}W_{i}^{ca}W_{i}^{db}\right)\right)\right)\label{eq:averaged_disorder}
\end{eqnarray}

Next, we introduce order parameters for the overlaps via delta functions
(we already have the one for the overlaps $\frac{1}{N}\sum W_{i}^{ca}\tilde{W}_{i}^{c}$
which must be equal to $S$), and use the expressions (\ref{eq:thetas_expansion}),
(\ref{eq:deltas_expansion}) and (\ref{eq:averaged_disorder}) in
eq.~(\ref{eq:omega1}), to get:
\begin{eqnarray}
\left\langle \Omega\left(S,y\right)^{n}\right\rangle _{\left\{ \xi^{\mu}\right\} } & = & \int\prod_{ic}d\mu\left(\tilde{W}_{i}^{c}\right)\int\prod_{ica}d\mu\left(W_{i}^{ca}\right)\int\prod_{ca\mu}\frac{d\lambda_{\mu}^{ca}d\hat{\lambda}_{\mu}^{ca}}{2\pi}\prod_{\mu}e^{i\left(\sum_{ca}\lambda_{\mu}^{ca}\hat{\lambda}_{\mu}^{ca}\right)}\times\nonumber \\
 &  & \times\int\prod_{c,a>b}\left(dq^{ca,cb}N\right)\delta\left(Nq^{ca,cb}-\sum_{i}W_{i}^{ca}W_{i}^{cb}\right)\times\nonumber \\
 &  & \times\int\prod_{c>d,ab}\left(dq^{ca,db}N\right)\delta\left(Nq^{ca,db}-\sum_{i}W_{i}^{ca}W_{i}^{db}\right)\times\nonumber \\
 &  & \times\prod_{\begin{array}{c}
ca\end{array}}\delta\left(NS-\sum_{i}W_{i}^{ca}\tilde{W}_{i}^{c}\right)\prod_{ca\mu}\Theta\left(\lambda_{\mu}^{ca}\right)\prod_{\mu}\exp\left(-\frac{1}{2}\sum_{ca}\left(\hat{\lambda}_{\mu}^{ca}\right)^{2}\right)\times\nonumber \\
 &  & \times\prod_{\mu}\exp\left(-\sum_{c}\sum_{a>b}\hat{\lambda}_{\mu}^{ca}\hat{\lambda}_{\mu}^{cb}q^{ca,cb}-\sum_{c>d}\sum_{ab}\hat{\lambda}_{\mu}^{ca}\hat{\lambda}_{\mu}^{db}q^{ca,db}\right)\label{eq:omega2}
\end{eqnarray}
Then we expand the deltas in the usual way introducing conjugate parameters
$\hat{q}^{ca,db}$ and $\hat{S}^{ca}$, and rearrange the integrals
such that we can factorize over $\mu$ (and therefore drop the $\mu$
index entirely) and over $i$ (and drop that index as well). We obtain:
\begin{eqnarray}
\left\langle \Omega\left(S,y\right)^{n}\right\rangle _{\left\{ \xi^{\mu}\right\} } & = & \int\prod_{c,a>b}\!\left(\frac{dq^{ca,cb}d\hat{q}^{ca,cb}N}{2\pi}\right)\!\int\prod_{c>d,ab}\!\left(\frac{dq^{ca,db}d\hat{q}^{ca,db}N}{2\pi}\right)\!\int\prod_{ca}\!\left(\frac{d\hat{S}^{ca}N}{2\pi}\right)\nonumber \\
 &  & e^{-N\left(\sum_{c}\sum_{a>b}q^{ca,cb}\hat{q}^{ca,cb}+\sum_{c>d}\sum_{ab}q^{ca,db}\hat{q}^{ca,db}+\sum_{ca}S\hat{S}^{ca}\right)}G_{S}^{N}\,G_{E}^{\alpha N}\label{eq:omega3}
\end{eqnarray}
where $G_{S}$ and $G_{E}$ are the entropic and the energetic terms,
respectively:
\begin{eqnarray}
G_{S} & = & \int\prod_{c}d\mu\left(\tilde{W}^{c}\right)\int\prod_{ca}d\mu\left(W^{ca}\right)\exp\left(\sum_{c}\sum_{a>b}\hat{q}^{ca,cb}W^{ca}W^{cb}+\right.\label{eq:Gs}\\
 &  & \quad\left.+\sum_{c>d}\sum_{ab}\hat{q}^{ca,db}W^{ca}W^{db}+\sum_{ca}\hat{S}^{ca}W^{ca}\tilde{W}^{c}\right)\nonumber 
\end{eqnarray}
\begin{eqnarray}
G_{E} & = & \int\prod_{ca}\frac{d\lambda^{ca}d\hat{\lambda}^{ca}}{2\pi}e^{i\left(\sum_{ca}\lambda^{ca}\hat{\lambda}^{ca}\right)}\prod_{ca}\Theta\left(\lambda^{ca}\right)\exp\left(-\frac{1}{2}\sum_{ca}\left(\hat{\lambda}^{ca}\right)^{2}+\right.\label{eq:Ge}\\
 &  & \quad\left.-\sum_{c}\sum_{a>b}\hat{\lambda}^{ca}\hat{\lambda}^{cb}q^{ca,cb}-\sum_{c>d}\sum_{ab}\hat{\lambda}^{ca}\hat{\lambda}^{db}q^{ca,db}\right)\nonumber 
\end{eqnarray}

\subsubsection{The external $1$-RSB Ansatz}

As explained in the main text, we will make a $1$-RSB Ansatz for
the planted configurations. More specifically, we will divide the
$n$ replicas in $\frac{n}{m}$ groups of $m$ replicas each, with
$m$ the Parisi $1$-RSB parameter over which we will subsequently
optimize. Let us then introduce the multi-index $c=\left(\alpha,\beta\right)$,
where $\alpha\in\left\{ 1,...,n/m\right\} $ labels a block of $m$
replicas, and $\beta\in\left\{ 1,...,m\right\} $ is the index of
replicas inside the block. This induces the following structure for
the overlap matrix $q^{\alpha\beta,a;\alpha^{\prime}\beta^{\prime},b}\equiv q^{ca,db}$:
\begin{equation}
q^{\alpha\beta,a;\alpha^{\prime}\beta^{\prime},b}=\begin{cases}
1 & \textrm{if}\,\alpha=\alpha^{\prime},\beta=\beta^{\prime},a=b\\
q_{2} & \textrm{if}\,\alpha=\alpha^{\prime},\beta=\beta^{\prime},a\ne b\\
q_{1} & \textrm{if}\,\alpha=\alpha^{\prime},\beta\ne\beta^{\prime}\\
q_{0} & \textrm{if}\,\alpha\ne\alpha^{\prime}
\end{cases}\label{eq:1rsb_ansatz}
\end{equation}
The structure of the conjugated parameters matrix $\hat{q}^{ca,db}$
is analogous. We also assume $\hat{S}^{ca}=\hat{S}$. Note that $\hat{S}$,
being the conjugate of $S$, takes the role of the soft constraint
parameter $\gamma$ of eq.~\ref{eq:local_num_sol_soft} that is used
throughout the main text. In fact, as already noted, studying the
soft-constrained system of eq.~\ref{eq:FreeEnergy} gives exactly
the same results with $\hat{S}=\gamma$, provided one makes an equivalent
symmetric Ansatz on the overlap $S$ as it is done for $\hat{S}$
here.

\subsubsection{Entropic term}

Let us consider the entropic term in the $1$-RSB Ansatz:
\begin{eqnarray}
G_{S} & = & \int\prod_{\alpha\beta}d\mu\left(\tilde{W}^{\alpha\beta}\right)\int\prod_{\alpha\beta,a}d\mu\left(W^{\alpha\beta,a}\right)\exp\left(-\frac{\hat{q}_{2}}{2}ny+\frac{\left(\hat{q}_{2}-\hat{q}_{1}\right)}{2}\sum_{\alpha\beta}\left(\sum_{a}W^{\alpha\beta,a}\right)^{2}\right)\times\nonumber \\
 &  & \times\exp\left(\frac{\left(\hat{q}_{1}-\hat{q}_{0}\right)}{2}\sum_{\alpha}\left(\sum_{\beta,a}W^{\alpha\beta,a}\right)^{2}+\frac{\hat{q}_{0}}{2}\left(\sum_{\alpha\beta,a}W^{\alpha\beta,a}\right)^{2}+\hat{S}\sum_{\alpha\beta,a}\tilde{W}^{\alpha\beta}W^{\alpha\beta,a}\right)
\end{eqnarray}
By means of a Hubbard-Stratonovich transformation
\[
\exp\left(\frac{b}{2}x^{2}\right)=\int Dz\exp\left(x\sqrt{b}z\right)
\]
on the term quadratic term $\left(\sum_{\alpha\beta,a}W^{\alpha\beta,a}\right)^{2}$,
everything factorizes over the replica index $\alpha$ , thus obtaining:
\begin{eqnarray}
G_{S} & = & e^{-\frac{\hat{q}_{2}}{2}ny}\int\!\!Dz_{0}\left[\prod_{\beta}d\mu\left(\tilde{W}^{\beta}\right)\int\prod_{\beta,a}d\mu\left(W^{\beta,a}\right)\exp\left(\frac{\left(\hat{q}_{2}-\hat{q}_{1}\right)}{2}\sum_{\beta}\left(\sum_{a}W^{\beta,a}\right)^{2}\right)\times\right.\nonumber \\
 &  & \left.\times\exp\left(\frac{\left(\hat{q}_{1}-\hat{q}_{0}\right)}{2}\left(\sum_{\beta,a}W^{\beta,a}\right)^{2}+z_{0}\sqrt{\hat{q}_{0}}\sum_{\beta,a}W^{\beta,a}+\hat{S}\sum_{\beta,a}\tilde{W}^{\beta}W^{\beta,a}\right)\right]^{\frac{n}{m}}
\end{eqnarray}
Another transformation of the term $\left(\sum_{\beta,a}W^{\alpha\beta,a}\right)$
allows to factorize over the index $\beta$:
\begin{eqnarray}
G_{S} & = & e^{-\frac{\hat{q}_{2}}{2}ny}\int\!\!Dz_{0}\left\{ \int\!\!Dz_{1}\left[\int\!\!d\mu\left(\tilde{W}\right)\int\prod_{a}d\mu\left(W^{a}\right)\exp\left(\frac{\left(\hat{q}_{2}-\hat{q}_{1}\right)}{2}\left(\sum_{a}W^{a}\right)^{2}\right)\times\right.\right.\nonumber \\
 &  & \left.\left.\times\exp\left(z_{1}\sqrt{\hat{q}_{1}-\hat{q}_{0}}\sum_{a}W^{a}+z_{0}\sqrt{\hat{q}_{0}}\sum_{a}W^{a}+\hat{S}\sum_{a}\tilde{W}W^{a}\right)\right]^{m}\right\} ^{\frac{n}{m}}
\end{eqnarray}
and again on the term $\left(\sum_{a}W^{a}\right)^{2}$, with a final
factorization over the index $a$:
\begin{eqnarray}
G_{S} & = & e^{-\frac{\hat{q}_{2}}{2}ny}\int\!\!Dz_{0}\left\{ \int\!\!Dz_{1}\left[\int\!\!Dz_{2}\int\!\!d\mu\left(\tilde{W}\right)\left(\int d\mu\left(W\right)\exp\left(z_{2}\sqrt{\hat{q}_{2}-\hat{q}_{1}}W\right)\times\right.\right.\right.\nonumber \\
 &  & \left.\left.\left.\times\exp\left(z_{1}\sqrt{\hat{q}_{1}-\hat{q}_{0}}W+z_{0}\sqrt{\hat{q}_{0}}W+\hat{S}\tilde{W}W\right)\right)^{y}\right]^{m}\right\} ^{\frac{m}{n}}
\end{eqnarray}

Let us then consider the specific case of binary variables $W,\tilde{W}\in\left\{ -1,+1\right\} $
and perform the sum over $W$ explicitly, thus obtaining:
\begin{equation}
G_{S}=e^{-\frac{\hat{q}_{2}}{2}ny}\int\!\!Dz_{0}\left\{ \int\!\!Dz_{1}\left[\int\!\!Dz_{2}\sum_{\tilde{W}=\pm1}\left(2\cosh\left(\tilde{A}\left(z_{0},z_{1},z_{2};\tilde{W}\right)\right)\right)^{y}\right]^{m}\right\} ^{\frac{n}{m}}
\end{equation}
where
\begin{equation}
\tilde{A}\left(z_{0},z_{1},z_{2};\tilde{W}\right)=z_{2}\sqrt{\hat{q}_{2}-\hat{q}_{1}}+z_{1}\sqrt{\hat{q}_{1}-\hat{q}_{0}}+z_{0}\sqrt{\hat{q}_{0}}+\hat{S}\tilde{W}
\end{equation}
Performing the limit $n\to0$ we obtain:
\begin{equation}
\frac{\log G_{S}}{n}=-\frac{\hat{q}_{2}}{2}y+\mathcal{G}_{S}
\end{equation}
where
\begin{equation}
\mathcal{G}_{S}=\frac{1}{m}\int\!\!Dz_{0}\log\left(\int\!\!Dz_{1}\left[\int\!\!Dz_{2}\sum_{\tilde{W}=\pm1}\left(2\cosh\left(\tilde{A}\left(z_{0},z_{1},z_{2};\tilde{W}\right)\right)\right)^{y}\right]^{m}\right)\label{eq:gs_1rsb_final}
\end{equation}

\subsubsection{Energetic term}

Let us plug the $1$-RSB Ansatz (\ref{eq:1rsb_ansatz}) in eq. (\ref{eq:Ge})
and get:
\begin{eqnarray}
G_{E} & = & \int\prod_{\alpha\beta,a}\frac{d\lambda^{\alpha\beta,a}d\hat{\lambda}^{\alpha\beta,a}}{2\pi}\prod_{\alpha\beta,a}\Theta\left(\lambda^{\alpha\beta,a}\right)\exp\left(i\sum_{\alpha\beta,a}\lambda^{\alpha\beta,a}\hat{\lambda}^{\alpha\beta,a}-\frac{1}{2}\sum_{\alpha\beta,a}\left(\hat{\lambda}^{\alpha\beta,a}\right)^{2}\right)\times\nonumber \\
 &  & \times\exp\left(-q_{2}\sum_{\alpha\beta}\sum_{a>b}\hat{\lambda}^{\alpha\beta,a}\hat{\lambda}^{\alpha\beta,b}-q_{1}\sum_{\alpha,\beta>\beta'}\sum_{ab}\hat{\lambda}^{\alpha\beta,a}\hat{\lambda}^{\alpha\beta^{\prime},b}\right)\times\nonumber \\
 &  & \times\exp\left(-q_{0}\sum_{\alpha>\alpha^{\prime},\beta\beta^{\prime}}\sum_{ab}\hat{\lambda}^{\alpha\beta,a}\hat{\lambda}^{\alpha^{\prime}\beta^{\prime},b}\right)\label{eq:ge_rsb_1}
\end{eqnarray}
We then use the formula
\[
\sum_{i>j}a_{i}a_{j}=\frac{1}{2}\left(\left(\sum_{i}a_{i}\right)^{2}-\sum_{i}a_{i}^{2}\right)
\]
for the various quadratic terms in $\lambda^{\prime}$s and $\hat{\lambda}^{\prime}$s
in eq. (\ref{eq:ge_rsb_1}), thus obtaining:

\begin{eqnarray}
G_{E} & = & \int\prod_{\alpha\beta,a}\frac{d\lambda^{\alpha\beta,a}d\hat{\lambda}^{\alpha\beta,a}}{2\pi}\prod_{\alpha\beta,a}\Theta\left(\lambda^{\alpha\beta,a}\right)\exp\left(i\sum_{\alpha\beta,a}\lambda^{\alpha\beta,a}\hat{\lambda}^{\alpha\beta,a}-\frac{1}{2}\sum_{\alpha\beta,a}\left(\hat{\lambda}^{\alpha\beta,a}\right)^{2}\right)\times\\
 &  & \times\exp\left(-q_{2}\sum_{\alpha\beta}\left(\left(\sum_{a}\hat{\lambda}^{\alpha\beta,a}\right)^{2}-\sum_{a}\left(\hat{\lambda}^{\alpha\beta,a}\right)^{2}\right)\right)\times\nonumber \\
 &  & \times\exp\left(-q_{1}\left(\sum_{\alpha}\left(\sum_{\beta,b}\hat{\lambda}^{\alpha\beta,a}\right)^{2}-\sum_{\beta,b}\left(\hat{\lambda}^{\alpha\beta,a}\right)^{2}\right)\right)\times\nonumber \\
 &  & \times\exp\left(-q_{0}\left(\left(\sum_{\alpha\beta,a}\hat{\lambda}^{\alpha\beta,a}\right)^{2}-\sum_{\alpha}\left(\sum_{\beta,a}\hat{\lambda}^{\alpha\beta,a}\right)^{2}\right)\right)\nonumber 
\end{eqnarray}

Let us then linearize the term multiplying $q_{0}$ by means of a
Hubbard-Stratonovich transformation, thereby factorizing over the
replica index $\alpha$:
\begin{eqnarray}
G_{E} & = & \int\!\!Dz_{0}\left[\prod\frac{d\lambda^{\beta,a}d\hat{\lambda}^{\beta,a}}{2\pi}\prod_{\beta,a}\Theta\left(\lambda^{\beta,a}\right)\exp\left(i\sum_{\beta,a}\lambda^{\beta,a}\hat{\lambda}^{\beta,a}-\frac{\left(1-q_{2}\right)}{2}\sum_{\beta,a}\left(\hat{\lambda}^{\beta,a}\right)^{2}\right)\times\right.\nonumber \\
 &  & \left.\times\exp\left(-\frac{\left(q_{2}-q_{1}\right)}{2}\sum_{\beta}\left(\sum_{a}\hat{\lambda}^{\beta,a}\right)^{2}-\frac{\left(q_{1}-q_{0}\right)}{2}\left(\sum_{\beta,a}\hat{\lambda}^{\beta,a}\right)^{2}\right)\times\right.\nonumber \\
 &  & \left.\times\exp\left(-iz_{0}\sqrt{q_{0}}\sum_{\beta,a}\hat{\lambda}^{\beta,a}\right)\right]^{\frac{n}{m}}
\end{eqnarray}
Performing two more Hubbard-Stratonovich transformations allows us
to factorize over the relevant indices $\beta$ and $a$:
\begin{equation}
G_{E}=\int\!\!Dz_{0}\left\{ \int\!\!Dz_{1}\left[\int\!\!Dz_{2}\,H\left(A\left(z_{0},z_{1},z_{2}\right)\right)^{y}\right]^{m}\right\} ^{\frac{n}{m}}
\end{equation}
where
\begin{equation}
A\left(z_{0},z_{1},z_{2}\right)=\frac{z_{0}\sqrt{q_{0}}+z_{1}\sqrt{q_{1}-q_{0}}+z_{2}\sqrt{q_{2}-q_{1}}}{\sqrt{1-q_{2}}}\label{eq:A_def}
\end{equation}
and we performed the Gaussian integral in $\hat{\lambda}$, writing
the definite Gaussian integral over $\lambda$ as an $H$ function,
$H\left(x\right)=\int_{x}^{+\infty}Dz$. In the limit $n\to0$ we
get:
\begin{equation}
\mathcal{G}_{E}=\frac{\log G_{E}}{n}=\frac{1}{m}\int\!\!Dz_{0}\log\left(\int\!\!Dz_{1}\left[\int\!\!Dz_{2}\,H\left(A\left(z_{0},z_{1},z_{2}\right)\right)^{y}\right]^{m}\right)\label{eq:ge_1rsb_final}
\end{equation}

\subsubsection{Final $1$-RSB expression}

Plugging eqs.~(\ref{eq:gs_1rsb_final}) and (\ref{eq:ge_1rsb_final})
into eq.~(\ref{eq:omega3}) we obtain:
\[
\left\langle \Omega\left(S,y\right)^{n}\right\rangle _{\left\{ \xi^{\mu}\right\} }=\exp\left(-Nny\mathscr{F}\left(S,y\right)\right)
\]
where we obtained an expression for eq.~(\ref{eq:F_def}):

\begin{equation}
\mathscr{F}\left(S,y\right)=-\left(y\frac{m}{2}q_{0}\hat{q}_{0}-y\frac{\left(m-1\right)}{2}q_{1}\hat{q}_{1}-\frac{\left(y-1\right)}{2}q_{2}\hat{q}_{2}-\frac{\hat{q}_{2}}{2}-S\hat{S}+\frac{1}{y}\mathcal{G}_{S}+\frac{\alpha}{y}\mathcal{G}_{E}\right)\label{eq:F_1RSB}
\end{equation}

The order parameters are obtained by the saddle point equations. In
order to study the zero-temperature limit $y\to\infty$, it is convenient
to rearrange the terms as:
\begin{equation}
\mathscr{F}\left(S,y\right)=-\left(\frac{my}{2}\left(q_{0}\hat{q}_{0}-q_{1}\hat{q}_{1}\right)-\frac{y}{2}\left(q_{2}\hat{q}_{2}-q_{1}\hat{q}_{1}\right)-\frac{\hat{q}_{2}}{2}\left(1-q_{2}\right)-S\hat{S}+\frac{1}{y}\mathcal{G}_{S}+\frac{\alpha}{y}\mathcal{G}_{E}\right)
\end{equation}
from which we see that in this limit the parameters must scale as:
\begin{eqnarray*}
m & \to & \frac{x}{y}\\
q_{2} & \to & q_{1}+\frac{\delta q}{y}\\
\hat{q}_{2} & \to & \hat{q}_{1}+\frac{\delta\hat{q}}{y}
\end{eqnarray*}
giving, to the leading order in $y$:

\begin{equation}
\mathscr{F}\left(S,\infty\right)=-\left(\frac{x}{2}\left(q_{0}\hat{q}_{0}-q_{1}\hat{q}_{1}\right)-\frac{1}{2}\left(q_{1}\delta\hat{q}+\hat{q}_{1}\delta q\right)-\frac{\hat{q}_{1}}{2}\left(1-q_{1}\right)-S\hat{S}+\mathcal{G}_{S}^{\infty}+\alpha\mathcal{G}_{E}^{\infty}\right)
\end{equation}
where
\begin{eqnarray}
\mathcal{G}_{S}^{\infty} & = & \frac{1}{x}\int\!\!Dz_{0}\log\left(\int\!\!Dz_{1}\,e^{x\,\tilde{B}\left(z_{0},z_{1}\right)}\right)\\
\tilde{B}\left(z_{0},z_{1}\right) & = & \max_{z_{2}\in\mathbb{R},\tilde{W}=\pm1}\left(\tilde{A}^{\infty}\left(z_{0},z_{1},z_{2};\tilde{W}\right)\right)\\
\tilde{A}^{\infty}\left(z_{0},z_{1},z_{2};\tilde{W}\right) & = & -\frac{z_{2}^{2}}{2}+\log\left(2\cosh\left(z_{2}\sqrt{\delta\hat{q}}+z_{1}\sqrt{\hat{q}_{1}-\hat{q}_{0}}+z_{0}\sqrt{\hat{q}_{0}}+\hat{S}\tilde{W}\right)\right)\\
\mathcal{G}_{E}^{\infty} & = & \frac{1}{x}\int\!\!Dz_{0}\log\left(\int\!\!Dz_{1}\,e^{x\,B\left(z_{0},z_{1}\right)}\right)\\
B\left(z_{0},z_{1}\right) & = & \max_{z_{2}}\left(A^{\infty}\left(z_{0},z_{1},z_{2}\right)\right)\\
A^{\infty}\left(z_{0},z_{1},z_{2}\right) & = & -\frac{z_{2}^{2}}{2}+\log\left(H\left(\frac{z_{0}\sqrt{q_{0}}+z_{1}\sqrt{q_{1}-q_{0}}+z_{2}\sqrt{\delta q}}{\sqrt{1-q_{1}}}\right)\right)
\end{eqnarray}

The expressions for $\mathcal{G}_{S}^{\infty}$ and $\mathcal{G}_{E}^{\infty}$
are obtained by the saddle point method, using $y\to\infty$.

The resulting saddle point equations for the order parameters are:
\begin{eqnarray}
\hat{q}_{0} & = & \frac{\alpha}{x\sqrt{\delta q}}\int\!\!Dz_{0}\frac{\int\!Dz_{1}\,e^{x\,B\left(z_{0},z_{1}\right)}z_{E}^{\star}\left(z_{0},z_{1}\right)\left(\frac{z_{1}}{\sqrt{q_{1}-q_{0}}}-\frac{z_{0}}{\sqrt{q_{0}}}\right)}{\int\!Dz_{1}\,e^{x\,B\left(z_{0},z_{1}\right)}}\\
\hat{q}_{1} & = & \frac{\alpha}{\delta q}\int\!\!Dz_{0}\frac{\int\!Dz_{1}\,e^{x\,B\left(z_{0},z_{1}\right)}\left(z_{E}^{\star}\left(z_{0},z_{1}\right)\right)^{2}}{\int\!Dz_{1}\,e^{x\,B\left(z_{0},z_{1}\right)}}\\
\delta\hat{q} & = & \left(1-x\right)\hat{q}_{1}+\frac{\alpha}{\sqrt{\delta q}}\int\!\!Dz_{0}\frac{\int\!Dz_{1}\,e^{x\,B\left(z_{0},z_{1}\right)}\left(z_{E}^{\star}\left(z_{0},z_{1}\right)\frac{z_{1}}{\sqrt{q_{1}-q_{0}}}+\frac{b\left(z_{0},z_{1}\right)}{\sqrt{\delta q}}\right)}{\int\!Dz_{1}\,e^{x\,B\left(z_{0},z_{1}\right)}}\\
q_{0} & = & \frac{1}{x\sqrt{\delta\hat{q}}}\int\!\!Dz_{0}\frac{\int\!Dz_{1}\,e^{x\,\tilde{B}\left(z_{0},z_{1}\right)}z_{S}^{\star}\left(z_{0},z_{1}\right)\left(\frac{z_{1}}{\sqrt{\hat{q}_{1}-\hat{q}_{0}}}-\frac{z_{0}}{\sqrt{\hat{q}_{0}}}\right)}{\int\!Dz_{1}\,e^{x\,\tilde{B}\left(z_{0},z_{1}\right)}}\\
q_{1} & = & \frac{1}{\delta\hat{q}}\int\!\!Dz_{0}\frac{\int\!Dz_{1}\,e^{x\,\tilde{B}\left(z_{0},z_{1}\right)}\left(z_{S}^{\star}\left(z_{0},z_{1}\right)\right)^{2}}{\int\!Dz_{1}\,e^{x\,\tilde{B}\left(z_{0},z_{1}\right)}}\\
\delta q & = & \left(1-x\right)q_{1}-1+\frac{1}{\sqrt{\delta\hat{q}}}\int\!\!Dz_{0}\frac{\int\!Dz_{1}\,e^{x\,\tilde{B}\left(z_{0},z_{1}\right)}z_{S}^{\star}\left(z_{0},z_{1}\right)\left(\frac{z_{1}}{\sqrt{\hat{q}_{1}-\hat{q}_{0}}}\right)}{\int\!Dz_{1}\,e^{x\,\tilde{B}\left(z_{0},z_{1}\right)}}\\
S & = & \frac{1}{\sqrt{\delta\hat{q}}}\int\!\!Dz_{0}\frac{\int\!Dz_{1}\,e^{x\,\tilde{B}\left(z_{0},z_{1}\right)}z_{S}^{\star}\left(z_{0},z_{1}\right)\mathrm{sign}\!\left(\hat{S}\left(z_{1}\sqrt{\hat{q}_{1}-\hat{q}_{0}}+z_{0}\sqrt{\hat{q}_{0}}\right)\right)}{\int\!Dz_{1}\,e^{x\,\tilde{B}\left(z_{0},z_{1}\right)}}
\end{eqnarray}
where
\begin{eqnarray}
z_{S}^{\star}\left(z_{0},z_{1}\right) & = & \argmax_{z_{2}\in\mathbb{R}}\left(\max_{\tilde{W}=\pm1}\left(\tilde{A}^{\infty}\left(z_{0},z_{1},z_{2};\tilde{W}\right)\right)\right)\\
z_{E}^{\star}\left(z_{0},z_{1}\right) & = & \argmax_{z_{2}\in\mathbb{R}}\left(A^{\infty}\left(z_{0},z_{1},z_{2}\right)\right)\\
b\left(z_{0},z_{1}\right) & = & \frac{z_{E}^{\star}\left(z_{0},z_{1}\right)\sqrt{\delta q}}{1-q_{1}}\left(z_{0}\sqrt{q_{0}}+z_{1}\sqrt{q_{1}-q_{0}}+z_{E}^{\star}\left(z_{0},z_{1}\right)\sqrt{\delta q}\right)
\end{eqnarray}

The parameter $x$ is implicitly set by the equation:
\begin{equation}
\frac{1}{2}\left(q_{0}\hat{q}_{0}-q_{1}\hat{q}_{1}\right)+\frac{\partial\mathcal{G}_{S}^{\infty}}{\partial x}+\alpha\frac{\partial\mathcal{G}_{E}^{\infty}}{\partial x}=0
\end{equation}
where
\begin{eqnarray}
\frac{\partial\mathcal{G}_{S}^{\infty}}{\partial x} & = & \frac{1}{x}\int\!\!Dz_{0}\frac{\int\!Dz_{1}\,e^{x\,\tilde{B}\left(z_{0},z_{1}\right)}\tilde{B}\left(z_{0},z_{1}\right)}{\int\!Dz_{1}\,e^{x\,\tilde{B}\left(z_{0},z_{1}\right)}}-\frac{\mathcal{G}_{S}^{\infty}}{x}\\
\frac{\partial\mathcal{G}_{E}^{\infty}}{\partial x} & = & \frac{1}{x}\int\!\!Dz_{0}\frac{\int\!Dz_{1}\,e^{x\,B\left(z_{0},z_{1}\right)}B\left(z_{0},z_{1}\right)}{\int\!Dz_{1}\,e^{x\,B\left(z_{0},z_{1}\right)}}-\frac{\mathcal{G}_{E}^{\infty}}{x}
\end{eqnarray}

In this limit, and since we optimize over $x$, $-\mathscr{F}\left(S,\infty\right)$
is equal to the \emph{local entropy} $\mathscr{S}_{I}$, i.e.~the
entropy of the solutions $W$ (which has formally the role of an energy
in our model). This is shown in Fig.~\ref{fig:Entropies}. Note that
the relationship between $\mathscr{F}$ and $\mathscr{S}_{I}$ is
different from the one reported in the main text, eq.~(\ref{eq:local_entropy}),
because here we have used $S$ directly as a control parameter rather
than $\gamma$ and thus no Legendre transform is required. The \emph{external
entropy}, i.e.~the entropy of the reference configurations $\tilde{W}$,
is the sum of two terms in the $1$-RSB scenario. The first, usually
called \emph{complexity}, accounts for the number of clusters of $\tilde{W}$,
and is implicitly set to zero by optimizing $\mathscr{F}$ over $x$
(see above). The second accounts for the number of configurations
in each cluster (it is usually just called entropy, but here we need
to qualify the name to avoid the confusion with the local entropy,
which in our model has formally the role of an energy), and is computed
from a first-order expansion in $y$, since $\mathscr{F}\left(S,y\right)=-\mathscr{S}_{I}-\frac{1}{y}\mathscr{S}_{E}$,
giving:
\begin{eqnarray}
\mathscr{S}_{E} & = & \frac{1}{2}\left(-\delta\hat{q}-\delta q\delta\hat{q}+\delta\hat{q}q_{1}+\delta q\hat{q}_{1}\right)+\mathcal{C}_{S}^{\infty}+\alpha\mathcal{C}_{E}^{\infty}\label{eq:ext-entropy}\\
\mathcal{C}_{S}^{\infty} & = & -\frac{1}{2}\int\!\!Dz_{0}\frac{\int\!Dz_{1}\,e^{x\,\tilde{B}\left(z_{0},z_{1}\right)}\log\left(1-\delta\hat{q}+z_{S}^{\star}\left(z_{0},z_{1}\right)^{2}\right)}{\int\!Dz_{1}\,e^{x\,\tilde{B}\left(z_{0},z_{1}\right)}}\\
\mathcal{C}_{E}^{\infty} & = & -\frac{1}{2}\int\!\!Dz_{0}\frac{\int\!Dz_{1}\,e^{x\,B\left(z_{0},z_{1}\right)}\left(\log\left(1+z_{E}^{\star}\left(z_{0},z_{1}\right)^{2}+b\left(z_{0},z_{1}\right)\right)-b\left(z_{0},z_{1}\right)\right)}{\int\!Dz_{1}\,e^{x\,B\left(z_{0},z_{1}\right)}}
\end{eqnarray}

In principle the external entropy should be non-negative since we
are dealing with a model with discrete variables. Numerical evaluation
of eq.~(\ref{eq:ext-entropy}) gives negative results, indicating
that further steps of replica symmetry breaking are needed. However,
the magnitude of the external entropy is very small ($\sim-10^{-6}$)
and vanishes as $S\to1$, therefore we expect that the corrections
of further RSB steps would be very small and that the correct value
of the external entropy would be $0$ (see also the discussion in
Sec.~\ref{sub:1RSB}). The physical meaning of this result is that
there is a sub-exponential number of configurations that can be defined
as the center of the cluster, i.e.~that yield the maximal local entropy
at a given overlap, and that they are arranged in a sub-exponential
number of distinct clusters.

\subsection{Reference configurations energy and constrained case\label{sub:Reference-configurations-energy}}

\subsubsection{Breaking the symmetry over reference configurations}

The expression of eq.~(\ref{eq:F_1RSB}) does not involve any overlaps
relative to the reference configurations. However, we wish to compute
the average energy associated with the reference configurations, which
will involve such quantities (see below). Therefore, we need to preliminarily
extract that information. To this end, we study a modified free energy
with respect to eq.~(\ref{eq:F_def}):
\begin{eqnarray}
\mathscr{F}_{C}\left(S,y\right) & = & -\frac{1}{Ny}\left\langle \log\left(\Omega_{C}\left(S,y\right)\right)\right\rangle _{\left\{ \xi^{\mu}\right\} }\label{eq:F_constrained_def}\\
 & = & -\frac{1}{Ny}\left\langle \log\left(\sum_{\left\{ \tilde{W}\right\} }\,f\left(\frac{1}{\sqrt{N}}\sum_{i}\tilde{W}_{i}\xi_{i}^{\mu}\right)\mathcal{\,N}_{\xi}\left(\tilde{W},S\right)^{y}\right)\right\rangle _{\left\{ \xi^{\mu}\right\} }\nonumber 
\end{eqnarray}
i.e.~one in which an additional term $f\left(\frac{1}{\sqrt{N}}\sum_{i}\tilde{W}_{i}\xi_{i}^{\mu}\right)$
was included in the expression. If we set the function $f\left(x\right)=\Theta\left(x\right)$
we get the case in which the reference solutions $\tilde{W}$ are
also constrained to be solutions to the classification problem. If
we set it to some function of a parameter $\eta$ such that $\lim_{\eta\to0}f\left(x\right)=1$,
we recover the previous case in the limit $\eta\to0$, i.e.~this
amounts to introduce a symmetry-breaking term and then making it vanish
at the end of the computation.

The computation follows along the lines of the previous case, but
requires the introduction of some additional order parameters: $\tilde{q}^{\alpha\beta;\alpha^{\prime}\beta^{\prime}}=\frac{1}{\sqrt{N}}\tilde{W}^{\alpha\beta}\cdot\tilde{W}^{\alpha^{\prime}\beta^{\prime}}$
for the overlaps between reference configurations and $S^{\alpha^{\prime}\beta^{\prime};\alpha\beta a}=\frac{1}{\sqrt{N}}\tilde{W}^{\alpha^{\prime}\beta^{\prime}}\cdot W^{\alpha\beta;a}$
for the overlaps between reference configurations and solutions. With
the $1$-RSB Ansatz, and including the constraint on the overlaps,
we have:

\begin{eqnarray}
\tilde{q}^{\alpha\beta;\alpha^{\prime}\beta^{\prime}} & = & \begin{cases}
1 & \textrm{if}\,\alpha=\alpha^{\prime},\beta=\beta^{\prime}\\
\tilde{q}_{1} & \textrm{if}\,\alpha=\alpha^{\prime},\beta\ne\beta^{\prime}\\
\tilde{q}_{0} & \textrm{if}\,\alpha\ne\alpha^{\prime}
\end{cases}\label{eq:1rsb_ansatz_constrained}\\
S^{\alpha^{\prime}\beta^{\prime};\alpha\beta a} & = & \begin{cases}
S & \textrm{if}\,\alpha=\alpha^{\prime},\beta=\beta^{\prime}\\
\tilde{S}_{1} & \textrm{if}\,\alpha=\alpha^{\prime},\beta\ne\beta^{\prime}\\
\tilde{S}_{0} & \textrm{if}\,\alpha\ne\alpha^{\prime}
\end{cases}\nonumber 
\end{eqnarray}
and analogous expressions for the conjugate parameters. The final
expression for the free energy is:
\begin{eqnarray}
\mathscr{F}_{C}\left(S,y\right) & = & -\left(\frac{m}{2y}\left(\tilde{q}_{0}\hat{\tilde{q}}_{0}-\tilde{q}_{1}\hat{\tilde{q}}_{1}\right)-\frac{1}{2y}\hat{\tilde{q}}_{1}\left(1-\tilde{q}_{1}\right)+\frac{my}{2}\left(q_{0}\hat{q}_{0}-q_{1}\hat{q}_{1}\right)+\right.\nonumber \\
 &  & \quad-\frac{y}{2}\left(q_{2}\hat{q}_{2}-q_{1}\hat{q}_{1}\right)-\frac{\hat{q}_{2}}{2}\left(1-q_{2}\right)-\left(S\hat{S}-\tilde{S}_{1}\hat{\tilde{S}}_{1}\right)+\nonumber \\
 &  & \quad\left.+m\left(\tilde{S}_{0}\hat{\tilde{S}}_{0}-\tilde{S}_{1}\hat{\tilde{S}}_{1}\right)+\frac{1}{y}\mathcal{G}_{CS}+\frac{\alpha}{y}\mathcal{G}_{CE}\right)
\end{eqnarray}
where
\begin{eqnarray}
\mathcal{G}_{CS} & = & \frac{1}{m}\int\!\!D\tilde{z}_{0}\int\!\!Dz_{0}\log\left(\int\!\!D\tilde{z}_{1}\int\!\!Dz_{1}\left[\sum_{\tilde{W}=\pm1}e^{\tilde{W}\,K\left(\tilde{z}_{0},z_{0},\tilde{z}_{1},z_{1}\right)}\:\times\right.\right.\\
 &  & \quad\times\left.\left.\int\!\!Dz_{2}\left(2\cosh\left(\tilde{A}_{C}\left(z_{0},z_{1},z_{2};\tilde{W}\right)\right)\right)^{y}\right]^{m}\right)\nonumber \\
\tilde{A}_{C}\left(z_{0},z_{1},z_{2};\tilde{W}\right) & = & z_{2}\sqrt{\hat{q}_{2}-\hat{q}_{1}}+z_{1}\sqrt{\hat{q}_{1}-\hat{q}_{0}}+z_{0}\sqrt{\hat{q}_{0}}+\left(\hat{S}-\hat{\tilde{S}}_{1}\right)\tilde{W}\\
K\left(\tilde{z}_{0},z_{0},\tilde{z}_{1},z_{1}\right) & = & \tilde{z}_{1}\sqrt{\left(\hat{\tilde{q}}_{1}-\hat{\tilde{q}}_{0}\right)-\frac{\left(\hat{\tilde{S}}_{1}-\hat{\tilde{S}}_{0}\right)^{2}}{\hat{q}_{1}-\hat{q}_{0}}}+z_{1}\frac{\hat{\tilde{S}}_{1}-\hat{\tilde{S}}_{0}}{\sqrt{\hat{q}_{1}-\hat{q}_{0}}}+\\
 &  & \quad+\tilde{z}_{0}\sqrt{\hat{\tilde{q}}_{0}-\frac{\left(\hat{\tilde{S}}_{0}\right)^{2}}{\hat{q}_{0}}}+z_{0}\frac{\hat{\tilde{S}}_{0}}{\sqrt{\hat{q}_{0}}}\nonumber 
\end{eqnarray}
\begin{eqnarray}
\mathcal{G}_{CE} & = & \frac{1}{m}\int\!\!D\tilde{z}_{0}\int\!\!Dz_{0}\log\left(\int\!\!D\tilde{z}_{1}\int\!\!Dz_{1}\left[\int\!\!Dz_{2}H\left(A\left(z_{0},z_{1},z_{2}\right)^{y}\right)\:\times\right.\right.\\
 &  & \quad\times\left.\left.L\left(\tilde{z}_{0},z_{0},\tilde{z}_{1},z_{1},z_{2}\right)\right]^{m}\right)\nonumber \\
L\left(\tilde{z}_{0},z_{0},\tilde{z}_{1},z_{1},z_{2}\right) & = & \int\!\!D\tilde{\lambda}\,f\left(\frac{z_{2}\frac{S-\tilde{S}_{1}}{\sqrt{1-\tilde{q}_{1}}}+\tilde{\lambda}\sqrt{\left(q_{2}-q_{1}\right)-\frac{\left(S-\tilde{S}_{1}\right)^{2}}{1-\tilde{q}_{1}}}}{\sqrt{q_{2}-q_{1}}}+z_{1}\frac{\tilde{S}_{1}-\tilde{S}_{0}}{\sqrt{q_{1}-q_{0}}}+\right.\\
 &  & \quad\left.+z_{0}\frac{\tilde{S}_{0}}{\sqrt{q_{0}}}+\tilde{z}_{1}\sqrt{\left(\tilde{q}_{1}-\tilde{q}_{0}\right)-\frac{\left(\tilde{S}_{1}-\tilde{S}_{0}\right)^{2}}{q_{1}-q_{0}}}+\tilde{z}_{0}\sqrt{\tilde{q}_{0}-\frac{\left(\tilde{S}_{0}\right)^{2}}{q_{0}}}\right)\nonumber 
\end{eqnarray}

From these expression, we can note that the dependency on the function
$f$ only enters the equations through the expression of $L$ in $\mathcal{G}_{CE}$.
Also, this expression does not depend on $y$. This has two consequences:
\begin{enumerate}
\item In the case where $\lim_{\eta\to0}f\left(x\right)=1$, $\mathcal{G}_{CE}\to\mathcal{G}_{E}$,
i.e.~the expression does not depend any more on $\tilde{q}_{1}$,
$\tilde{q}_{0}$, $\tilde{S}_{1}$ or $\tilde{S}_{0}$ and simplifies
to the previous case, as expected. In turn, this implies that the
conjugated order parameters $\hat{\tilde{q}}_{1},$ $\hat{\tilde{q}}_{0}$,
$\hat{\tilde{S}}_{1}$ and $\hat{\tilde{S}}_{0}$ all tend to $0$,
thus reducing the expression of the free energy to the previous case
eq.~(\ref{eq:F_1RSB});
\item In the limit $y\to\infty$ in the constrained case ($f\left(x\right)=\Theta\left(x\right)$),
we also have $\mathcal{G}_{CE}\to\mathcal{G}_{E}$, since the term
with $y$ in the exponent dominates the saddle point expansion. Again
we recover expression~(\ref{eq:F_1RSB}) for the free energy of the
system, which means that the local entropy is unchanged in the constrained
case. This may suggest that, even in the unconstrained case, the reference
configuration $\tilde{W}$ is never ``too far'' from an actual solution
to the problem (more precisely, within a distance $o\left(N\right)$
from a solution). Note, however, that --- as one would expect ---
the external entropy is different in this case, since it depends on
the first order expansion in $y$, which is affected by the $L$ term.
\end{enumerate}
Following observation 1, we can derive the expression for the order
parameters $\tilde{q}_{1}$, $\tilde{q}_{0}$, $\tilde{S}_{1}$ and
$\tilde{S}_{0}$ by using the saddle point equations and by assuming
that the conjugate parameters $\hat{\tilde{q}}_{1},$ $\hat{\tilde{q}}_{0}$,
$\hat{\tilde{S}}_{1}$ and $\hat{\tilde{S}}_{0}$ are of order $\eta\ll1$
and taking the leading order in the resulting expression. We actually
only need the results for $\tilde{S}_{1}$ and $\tilde{S}_{0}$, which
turn out to be:
\begin{eqnarray}
\tilde{S}_{1} & = & \frac{S}{1-m}+\frac{1}{y\left(m-1\right)\sqrt{\hat{q}_{1}-\hat{q}_{0}}}\int Dz_{0}\frac{I_{dz}\left(z_{0}\right)}{I_{s}\left(z_{0}\right)}\\
\tilde{S}_{0} & = & \frac{1}{my}\left(\frac{1}{\sqrt{\hat{q}_{1}-\hat{q}_{0}}}\int Dz_{0}\frac{I_{dz}\left(z_{0}\right)}{I_{s}\left(z_{0}\right)}-\frac{1}{\sqrt{\hat{q}_{0}}}\int Dz_{0}\,z_{0}\frac{I_{d}\left(z_{0}\right)}{I_{s}\left(z_{0}\right)}\right)\\
I_{s}\left(z_{0}\right) & = & \int\!\!Dz_{1}\left[\int Dz_{2}\sum_{\tilde{W}=\pm1}\left(2\cosh\left(\tilde{A}\left(z_{0},z_{1},z_{2};\tilde{W}\right)\right)\right)^{y}\right]^{m}\\
I_{d}\left(z_{0}\right) & = & \int\!\!Dz_{1}\left[\int Dz_{2}\sum_{\tilde{W}=\pm1}\left(2\cosh\left(\tilde{A}\left(z_{0},z_{1},z_{2};\tilde{W}\right)\right)\right)^{y}\right]^{m-1}\times\\
 &  & \quad\times\left(\int Dz_{2}\sum_{\tilde{W}=\pm1}\tilde{W}\left(2\cosh\left(\tilde{A}\left(z_{0},z_{1},z_{2};\tilde{W}\right)\right)\right)^{y}\right)\nonumber \\
I_{dz}\left(z_{0}\right) & = & \int\!\!Dz_{1}\,z_{1}\left[\int Dz_{2}\sum_{\tilde{W}=\pm1}\left(2\cosh\left(\tilde{A}\left(z_{0},z_{1},z_{2};\tilde{W}\right)\right)\right)^{y}\right]^{m-1}\times\\
 &  & \quad\times\left(\int Dz_{2}\sum_{\tilde{W}=\pm1}\tilde{W}\left(2\cosh\left(\tilde{A}\left(z_{0},z_{1},z_{2};\tilde{W}\right)\right)\right)^{y}\right)\nonumber 
\end{eqnarray}

In the limit $y\to\infty$, we have the scaling:
\begin{eqnarray*}
\tilde{S}_{1} & = & S-\frac{\delta S}{y}
\end{eqnarray*}
and finally the expressions:
\begin{eqnarray}
\delta S & = & \frac{1}{\sqrt{\hat{q}_{1}-\hat{q}_{0}}}\int\!\!Dz_{0}\frac{\int\!Dz_{1}\,z_{1}e^{x\,\tilde{B}\left(z_{0},z_{1}\right)}\tilde{W}^{\star}\left(z_{0},z_{1}\right)}{\int\!Dz_{1}e^{x\,\tilde{B}\left(z_{0},z_{1}\right)}}-xS\\
\tilde{S}_{0} & = & \frac{1}{x\sqrt{\hat{q}_{1}-\hat{q}_{0}}}\int\!\!Dz_{0}\frac{\int\!Dz_{1}\,z_{1}e^{x\,\tilde{B}\left(z_{0},z_{1}\right)}\tilde{W}^{\star}\left(z_{0},z_{1}\right)}{\int\!Dz_{1}e^{x\,\tilde{B}\left(z_{0},z_{1}\right)}}+\\
 &  & \quad-\frac{1}{x\sqrt{\hat{q}_{0}}}\int\!\!Dz_{0}\,z_{0}\frac{\int Dz_{1}\,e^{x\,\tilde{B}\left(z_{0},z_{1}\right)}\tilde{W}^{\star}\left(z_{0},z_{1}\right)}{\int Dz_{1}e^{x\,\tilde{B}\left(z_{0},z_{1}\right)}}\nonumber 
\end{eqnarray}
where
\begin{eqnarray}
\tilde{W}^{\star}\left(z_{0},z_{1}\right) & = & \argmax_{\tilde{W}=\pm1}\left(\max_{z_{2}\in\mathbb{R}}\left(\tilde{A}^{\infty}\left(z_{0},z_{1},z_{2};\tilde{W}\right)\right)\right)
\end{eqnarray}

\subsubsection{Energy density}

In order to compute the typical energy density of the unconstrained
reference configurations $\tilde{W}$, we need to evaluate the probability
of classifying incorrectly a pattern $\xi^{\star}$ drawn at random
from the training set. This probability can be obtained by calculating:

\begin{eqnarray}
P\left(\sigma^{\star}\ne1\right) & = & \left\langle \Theta\left(-\frac{1}{\sqrt{N}}\sum_{i}\tilde{W}_{i}\xi_{i}^{\star}\right)\right\rangle _{\tilde{W}}\label{eq:Prob_of_solution=00003Denergy}
\end{eqnarray}

where the average is defined over the weighted measure $d\mu_{W}\left(\tilde{W}\right)=d\mu\left(\tilde{W}\right)\mathcal{N}_{\xi}\left(\tilde{W},S\right)^{y}$.
Since:

\begin{eqnarray}
\left\langle \prod_{\mu}\Theta\left(-\frac{1}{\sqrt{N}}\sum_{i}\tilde{W}_{i}\xi_{i}^{\mu}\right)\right\rangle _{\tilde{W}} & = & \frac{\int\!d\mu_{W}\left(\tilde{W}\right)\prod_{\mu}\Theta\left(-\frac{1}{\sqrt{N}}\sum_{i}\tilde{W}_{i}\xi_{i}^{\mu}\right)}{\int\!d\mu_{W}\left(\tilde{W}\right)}\label{eq:averageRatio}
\end{eqnarray}

this calculation can be carried out straightforwardly by exploiting
the replica trick, i.e.~by rewriting the ratio in (\ref{eq:averageRatio})
as:

\begin{eqnarray}
 &  & \lim_{n\to0}\int\!\!d\mu_{W}\left(\tilde{W}\right)\Theta\left(-\frac{1}{\sqrt{N}}\sum_{i}\tilde{W}_{i}\xi_{i}^{\star}\right)\left(\int d\mu_{W}\left(\tilde{W}\right)\right)^{n-1}=\label{eq:replicaTrick}\\
 &  & =\lim_{n\to0}\int\prod_{c}d\mu_{W}\left(\tilde{W}^{c}\right)\Theta\left(-\frac{1}{\sqrt{N}}\sum_{i}\tilde{W}_{i}^{1}\xi_{i}^{\star}\right)\nonumber 
\end{eqnarray}

where we have introduced $n-1$ unconstrained replicas of the reference
solution, leaving out the replica index $1$ for the $\tilde{W}$-replica
coupled to the pattern $\xi^{\star}$ by the constraint. In this way
the quenched disorder can be averaged out, and in the $n\to0$ limit
one recovers the initial expression.

As noted in the previous section, when one extracts the overlaps referred
to the reference configurations by introducing vanishing constraints
(i.e. when $\eta\to0$), the conjugate parameters related to these
overlaps tend to vanish as well.

Therefore, if one organizes the calculation similarly to the previous
ones, it is easy to see that in (\ref{eq:replicaTrick}) the entropic
terms cancel out and the only non-zero contribution to the average
comes from the energetic part $G_{E}^{\prime}$, where:

\begin{eqnarray}
G_{E}^{\prime} & = & \int\frac{d\tilde{\lambda}^{1}d\hat{\tilde{\lambda}}^{1}}{2\pi}\prod_{\alpha\beta,a}\frac{d\lambda^{\alpha\beta,a}d\hat{\lambda}^{\alpha\beta,a}}{2\pi}\Theta\left(-\tilde{\lambda}^{1}\right)\prod_{\alpha\beta,a}\Theta\left(\lambda^{\alpha\beta,a}\right)\times\nonumber \\
 &  & \times\exp\left(i\,\left(\tilde{\lambda}^{1}\hat{\tilde{\lambda}}^{1}+\sum_{\alpha\beta,a}\lambda^{\alpha\beta,a}\hat{\lambda}^{\alpha\beta,a}\right)-\frac{1}{2}\left(\left(\hat{\tilde{\lambda}}^{1}\right)^{2}+\sum_{\alpha\beta,a}\left(\hat{\lambda}^{\alpha\beta,a}\right)^{2}\right)\right)\times\nonumber \\
 &  & \times\exp\left(-q_{2}\sum_{\alpha\beta}\sum_{a>b}\hat{\lambda}^{\alpha\beta,a}\hat{\lambda}^{\alpha\beta,b}-q_{1}\sum_{\alpha,\beta>\beta'}\sum_{ab}\hat{\lambda}^{\alpha\beta,a}\hat{\lambda}^{\alpha\beta^{\prime},b}\right)\times\nonumber \\
 &  & \times\exp\left(-q_{0}\sum_{\alpha>\alpha^{\prime},\beta\beta^{\prime}}\sum_{ab}\hat{\lambda}^{\alpha\beta,a}\hat{\lambda}^{\alpha^{\prime}\beta^{\prime},b}-\sum_{a}\hat{\tilde{\lambda}}^{1}\hat{\lambda}^{11,a}\left(S-\tilde{S}_{1}\right)\right)\times\nonumber \\
 &  & \times\exp\left(-\sum_{\beta}\sum_{a}\hat{\tilde{\lambda}}^{1}\hat{\lambda}^{1\beta,a}\left(\tilde{S}_{1}-\tilde{S}_{0}\right)-\sum_{a\beta}\sum_{a}\hat{\tilde{\lambda}}^{1}\hat{\lambda}^{\alpha\beta,a}\tilde{S}_{0}\right)\label{eq:ge_energy}
\end{eqnarray}

The final expression we obtain is the following:

\begin{eqnarray}
 &  & P\left(\sigma^{\star}\ne1\right)=\label{eq:energy_finalexpr}\\
 &  & =\int\!\!Dz_{0}\frac{\int\!Dz_{1}\left(\int\!Dz_{2}H\left(A\left(z_{0},z_{1},z_{2}\right)\right)^{y}\right)^{m-1}\int\!Dz_{2}H\left(A\left(z_{0},z_{1},z_{2}\right)\right)^{y}H\left(-C\left(z_{0},z_{1},z_{2}\right)\right)}{\int\!Dz_{1}\left(\int\!Dz_{2}H\left(A\left(z_{0},z_{1},z_{2}\right)\right)^{y}\right)^{m}}\nonumber 
\end{eqnarray}

with the definition (\ref{eq:A_def}), and where:

\begin{eqnarray*}
C\left(z_{0},z_{1},z_{2}\right) & = & \frac{z_{0}\frac{\tilde{S}_{0}}{\sqrt{q_{0}}}+z_{1}\frac{\tilde{S}_{1}-\tilde{S}_{0}}{\sqrt{q_{1}-q_{0}}}+z_{2}\frac{S-\tilde{S}_{1}}{\sqrt{q_{2}-q_{1}}}}{\sqrt{1-\frac{\tilde{S}_{0}^{2}}{q_{0}}-\frac{\left(\tilde{S}_{1}-\tilde{S}_{0}\right)^{2}}{q_{1}-q_{0}}-\frac{\left(S-\tilde{S}_{1}\right)^{2}}{q_{2}-q_{1}}}}
\end{eqnarray*}

In the limit $y\to\infty$, we have:
\begin{eqnarray}
P\left(\sigma^{\star}\ne1\right) & = & \int\!\!Dz_{0}\frac{\int\!Dz_{1}e^{x\,B\left(z_{0},z_{1}\right)}H\left(-\frac{z_{0}\frac{\tilde{S}_{0}}{\sqrt{q_{0}}}+z_{1}\frac{\tilde{S}_{1}-\tilde{S}_{0}}{\sqrt{q_{1}-q_{0}}}+z_{2}\frac{\delta S}{\sqrt{\delta q}}}{\sqrt{1-\frac{\tilde{S}_{0}^{2}}{q_{0}}-\frac{\left(\tilde{S}_{1}-\tilde{S}_{0}\right)^{2}}{q_{1}-q_{0}}}}\right)}{\int\!Dz_{1}e^{x\,B\left(z_{0},z_{1}\right)}}
\end{eqnarray}
which is shown in Fig.~\ref{fig:Errors}

\end{document}